\documentclass[11pt]{article}
\usepackage{hyperref}
\usepackage{amssymb}
\usepackage{graphicx}
\usepackage{float}
\usepackage{slashed}
\usepackage{amsmath}
\usepackage{tabularx}
\usepackage{breqn}
\usepackage{authblk}
\usepackage{array}
\usepackage{verbatim}
\usepackage{dsfont}
\usepackage{cite}
\usepackage{physics}
\usepackage[section]{placeins}
\usepackage[a4paper, total={6.8in, 10in}]{geometry}
\numberwithin{equation}{section}
\newcommand{\be}{\begin{equation}}
\newcommand{\ee}{\end{equation}}
\newcommand{\bea}{\begin{eqnarray}}
\newcommand{\eea}{\end{eqnarray}}
\newcommand{\ba}{\begin{aligned}}
\newcommand{\ea}{\end{aligned}}

\begin{document}
\title{How does Casimir energy fall in $\kappa$-deformed space-time?}

\author[1]{E. Harikumar\thanks{eharikumar@uohyd.ac.in}} 
\affil[1]{School of Physics, University of Hyderabad, Central University P.O. Hyderabad-500046, Telangana, India}
\author[2]{K. V. Shajesh\thanks{kvshajesh@gmail.com}}
\affil[2]{School of Physics and Applied Physics, Southern Illinois University-Carbondale, Carbondale, Illinois 62901 USA}
\author[1]{Suman Kumar Panja\thanks{sumanpanja19@gmail.com}} 
%\affil[1]{School of Physics, University of Hyderabad, \\Central University P.O, Hyderabad-500046, Telangana, India}

\maketitle
\begin{abstract}
We investigate the response of Casimir energies to fluctuations in a scalar field in a weak gravitational field in the $\kappa$-deformed space-time. We model the Casimir plates in a gravitational field by $\kappa$-deformed Rindler coordinates and calculate the Casimir energy using the $\kappa$-deformed scalar field. We show that the Casimir energy accelerates in a weak gravitational field like a mass. Thus, our calculations show that the mass-energy equivalence principle holds in $\kappa$-deformed space-time even though a length scale is introduced through space-time non-commutativity.
\end{abstract}

\section{Introduction} 
A complete description of quantum gravity remains an open problem. General relativity describes the structure of space-time geometry at a large scale. In contrast, quantum mechanics describes the behavior of physical systems at a small scale. Constructing a consistent quantum theory of gravity is an active area of research in theoretical physics. This involves reconciling general relativity with principles of quantum mechanics. In quest of such quantum gravity theory, various approaches have been developed and studied vigorously \cite{seiberg-witten, Glikman,rov, am,madore,bal,connes, dop}. Existence of a fundamental length scale is a feature exhibited by all these approaches. Since models built on non-commutative space-times introduce a length scale, the relevance of such space-times to quantum gravity is being investigated thoroughly \cite{Glikman,connes, dop}. Non-commutative space-time also appears in the low energy limit of certain string theory models \cite{nek} and quantum gravity models \cite{livin}.

One of the initial motivations behind introducing non-commutative space-time was to remove divergences from QFT. Two different types of non-commutative space-times\cite{connes} that have been studied extensively are Moyal space-time\cite{nek} and $\kappa$-deformed space-time\cite{luke}. In the former, the coordinates satisfy $[\hat{x}^{\mu},\hat{x}^{\nu}]=i\theta^{\mu\nu}$, where $\theta^{\mu\nu}$ is an antisymmetric constant tensor with dimension of $(\text{length})^2$. It has been shown that this constant $\theta^{\mu\nu}$ leads to the violation of Lorentz invariance \cite{nek}. The deformed special relativistic transformations leave physics invariant in the $\kappa$-deformed space-time \cite{dsr,majid}. $\kappa$-deformed space-time is also shown to be associated with the low energy limit of certain quantum gravity models \cite{jerzy}, and its coordinates satisfy the following commutation relations
\begin{equation}
 [{\hat x}^i, {\hat x}^j]=0,~~ [{\hat x}^0,{\hat x}^i]=i a x^i,~~ a=\frac{1}{\kappa},\label{intro2}
\end{equation} 
where $a$  is the deformation parameter with the dimension of length. 

The existence of a fundamental length scale in quantum gravity models requires the modification of the principle of relativity. This led to the formulation of deformed special relativity\cite{dsr}. The space-time associated with deformed special relativity is $\kappa$-deformed space-time\cite{majid}. Non-commutative space-times have inherent features such as non-locality, non-linearity, and introduction of a length parameter. Symmetries corresponding to these space-times are realized as Hopf algebras\cite{Chaichan, Wess,luk,hopf}. Field theory models on such space-times have been constructed and studied with various motivations\cite{nek, luke}. Various models on different non-commutative space-times have been studied extensively to analyze the effect of the fundamental length scale and its possible signals\cite{jabbari,sdas,gupta,siva,kapoor,ravi,hari,btz,vish}. Thus, it is of intrinsic interest to study the implications of a minimal length (introduced through the space-time non-commutativity) in physical phenomena. One well-studied physical phenomenon where length scale plays an important role is the Casimir effect\cite{hbg, Mil,kam}. It is known that two parallel conducting plates separated by a distance attract each other. This force is known to arise due to the vacuum fluctuations of the quantized fields. 

Casimir effect has been studied in non-commutative space-times such as Moyal space-time and $\kappa$-deformed space-time \cite{casadio,fosco, pinto,skp} using scalar field theory. Using a coherent state approach and smeared boundary condition Casimir force has been calculated between two parallel conducting plates in Moyal space-time \cite{casadio,fosco}. In Refs.\cite{pinto,skp} the Casimir effect is investigated for massless scalar field using the Green's function approach and a bound on the non-commutative length scale was obtained \cite{skp}.
Correction to Casimir force and energy has been obtained in Ref.\cite{skp2} in the Doplicher-Fredenhagen-Roberts space-time \cite{dop}.

It has been shown in \cite{shajesh4,shajesh1,
shajesh2a,shajesh2b,shajesh3,saharian,caldwell} that Casimir energies gravitating in
a weak gravitational field respects the equivalence principle. For
earlier related work, see Refs.\cite{caldwell, sorge1,sorge2, lima,sorge3,calloni,bimonte,esposito} and discussion in \cite{shajesh1}.
Here, we investigate how this result fares in the $\kappa$-deformed
non-commutative space-time. This is of interest because introducing a fundamental length scale in these theories is known to modify the
principle of relativity. Similarly, it is of interest to inquire if such
length scales disturb the equivalence principle. Here, we show that in a
weak gravitational field associated with the proper acceleration of
the $\kappa$-deformed Rindler metric, the Casimir energy gravitates like masses. That is, the fundamental length scale of the $\kappa$-deformed non-commutative
space-time does not seem to disturb the equivalence principle in this
case, even though it alters the Casimir energy of the parallel plate
configuration. 

In Refs.\cite{shajesh1,shajesh2a,shajesh2b,saharian,shajesh3,shajesh4} it is shown that the Casimir energy accelerates. The gravitational force on a Casimir apparatus is calculated in weak gravitational field approximation for Fermi coordinates \cite{misner} and for isotropic Schwarzschild coordinates \cite{shajesh1}. In \cite{shajesh2a,shajesh2b}, authors found the gravitational force on two parallel plates in Rindler space-time \cite{rindler} using Green's function method. They showed that in the limit of the weak gravitational field, (that is in the limit of large Rindler coordinate) total energy, including the Casimir energy associated with the plates, fall. The same study has been done using the zeta-function regularization method, confirming the result in \cite{saharian}. In \cite{shajesh3}, the gravitational force acting on the two rotating parallel plates is discussed. Thus, in \cite{shajesh1,shajesh2a,shajesh2b,saharian}, it is shown that Casimir energy falls like gravitational mass in a weak gravitational field. In this paper, we calculate the Casimir energy in the $\kappa$-deformed Rindler coordinates and analyze the status of the equivalence principle. $\kappa$-deformed space-time being the underlying space-time of deformed/doubly special relativity, it is natural to ask whether the mass-energy equivalence principle still holds or gets modified if not violated. Our investigation also sheds light into the role of the fundamental length scale associated with quantum gravity in modifying the Casimir energy of parallel plate configuration in the deformed space-time.

In this study, we follow the procedure used in \cite{shajesh2a,shajesh2b} to investigate the gravitational force on the Casimir apparatus for both single plate and two parallel plates in the $\kappa$-deformed Rindler space-time. We model the semitransparent plates with $\delta$-function potentials in both cases. For single plate we consider plate at the Rindler coordinate $\xi=\xi_{1}$ representing proper acceleration  $\frac{1}{\xi_{1}}$. For two parallel plates, we consider the situation where the plates move rigidly in such a way that the proper distance between plates remains unchanged. This is achieved by considering that the plates move with different but constant proper acceleration. Here we consider two parallel plates at $\xi=\xi_{1}$ and $\xi=\xi_{2}$ in the Rindler cordinates, moving with proper accelerations $\frac{1}{\xi_{1}}$ and $\frac{1}{\xi_{2}}$, respectively. Next, we calculate the net force acting on a single plate and two parallel plates using Green's function method. For this, we first find the reduced Green function by solving the equation of motion for the system containing a single plate in the $\kappa$-deformed Rindler space-time. Then, we obtain force density using the energy-momentum tensor. By taking the vacuum expectation value of this force density expression and using jump discontinuity of reduced Green's function, we find the $\kappa$-derfomed net force on the single plate in the Minkowski-space limit (weak field approximation)\footnote{The Minkowski-space limit refers to the case where the gravitational field becomes extremely weak. In this limit, the space-time approaches the flat Minkowski space-time. The weak field approximation is a technique used in general relativity to simplify the equations in situations where gravitational fields are relatively weak. The weak field approximation converges to the Minkowski-space limit when gravitational effects become very small.}. This modified force has divergent contributions as in the commutative space-time \cite{shajesh2a,shajesh2b}. We derive the Casimir energy by repeating the same process for two parallel plates and we find the total energy associated with the two parallel plates in the $\kappa$-deformed Minkowski space-time. We show that both the divergent parts of the Casimir energy, coming from the self-energy of the plates, and the finite part of the Casimir energy, accelerates according to the equivalence principle even in the $\kappa$-deformed space-time. 

The organization of this paper is as follows. The next Section briefly reviews $\kappa$-deformed space-time and its symmetry algebra. In Section 3, we choose a specific realization for $\kappa$-deformed space-time. This allows us to write the non-commutative variables in terms of commutative variables, their derivatives, and the deformation parameter $a$. Using this in the generalized commutation relation for the phase space coordinates we obtain the $\kappa$-deformed Rindler metric. We construct $\kappa$-deformed Lagrangians describing interaction of a scalar field with a single plate and two parallel plates in $\kappa$-deformed Rindler space-time in Section 4. These Lagrangians, illustrating the interaction of $\kappa$-scalar field with either a single plate or two parallel plates are written in terms of commutative variables and are valid up to the first non-vanishing terms in the deformation parameter $a$. We find the equations of motion and symmetrized energy-momentum tensors using these Lagrangians. In Section 5, we perturbatively solve the equation of motion obtained in Section 4 for both single-plate interactions and two parallel-plate interactions. In both cases, we find modified reduced Green's function for different regions. We also take weak field approximation (Minkowski-space limit) of these solutions and find reduced Minkowski Green's function for different regions of interest, for both single plate and two parallel plates. Our main result regarding validity of the equivalence principle in a space-time with a fundamental length scale is obtained in Section 6. Here, we find energy density expression using components of the energy-momentum tensor. Next, we discuss the vacuum expectation value of force density in terms of Green's function. Using this, we find the total force on the single plate and two parallel plates separately in the Minkowski-space limit. Using this, we show that the Casimir energy falls freely in a weak gravitational field, showing that the mass-energy equivalence principle is unaffected by the non-commutativity of the space-time. We note that the $\kappa$-deformed Casimir energy is dependent on the fundamental length scale ``$a$". This indicates that the existance of a fudamental length scale (which is a signal of quantum gravity) does not violate equivalance principle. Our concluding remarks are given in Section 7. The appendix discusses the calculation of the total energy associated with the two parallel plates in $\kappa$-deformed  Minkowski space-time.

\section{$\kappa$-deformed space-time, symmetries, and dispersion relation}
This Section summarises the essential results related to the realization of the $\kappa$-deformed space-time coordinates and the corresponding symmetry algebra \cite{hopf}. 

$\kappa$-deformed space-time is a Lie-algebraic non-commutative space-time, coordinates satisfy Eq.(\ref{intro2}). Using the star product formalism, field theory models on the $\kappa$ space-time were built\cite{dimitrijevic,dasz}. This formalism replaces the usual notion of the pointwise product between the coordinates (and their functions) with the star product, which is invariant under the $\kappa$-Poincare algebra \cite{dimitrijevic,dasz}. Alternatively, one can study the field theory models using the approach where the non-commutative coordinates are represented in terms of the functions of commutative coordinates and their derivatives \cite{hopf,mel2}. This approach has been shown to be equivalent to the star product formalism in Ref.\cite{mel3}. In this work, we will use the realization method \cite{hopf,mel2,mel3} and thus the coordinates of $\kappa$-deformed space-times are represented in terms of commutative coordinates $x_{\mu}$ and their derivatives  as \cite{hopf}
\begin{equation}\label{CFK2}
\begin{split}
 \hat{x}_0=&x_0\psi(A)+iax_j\partial_j\gamma(A),\\
 \hat{x}_i=&x_i\varphi(A),
\end{split}
\end{equation}
where $A=ia\partial_0=ap^{0}$ and $\psi$, $\gamma$ and $\varphi$ are functions of $A$ satisfying conditions
\begin{equation}\label{CFK3}
 \psi(0)=1,~\varphi(0)=1.
\end{equation}   
We substitute Eq.(\ref{CFK2}) in Eq.(\ref{intro2}) and find
\begin{equation}\label{CFK4}
 \frac{\varphi'(A)}{\varphi(A)}\psi(A)=\gamma(A)-1,
\end{equation} 
where $\varphi^{\prime}=\frac{d\varphi}{dA}$. Two possible realisations of $\psi(A)$ are $\psi(A)=1$ and $\psi(A)=1+2A$ \cite{hopf}. Here onwards, we choose $\psi(A)=1$. 
Allowed choices of $\varphi$ are $e^{-A}, e^{-\frac{A}{2}}, 1, \frac{A}{e^A-1}$, etc., and different choices of $\varphi$ link to various ordering choices \cite{hopf}. In this study we  use the $\varphi=e^{-\frac{A}{2}}$ realization. Thus, from Eq.(\ref{CFK2}) we find 
\be
\hat{x}_0=x_0~~\text{and}~~ \hat{x}_i=x_i e^{-\frac{A}{2}}. \label{CFK6a}
\ee 
Symmetry algebra of the $\kappa$-Minkowski space-time is the $\kappa$-Poincare algebra \cite{majid,luk1,luk2}. Thus, the commutation relations of the Poincare algebra gets modified. Alternatively, the symmetry algebra can be realized with the standard Poincare algebra, but the generators of the algebra are modified. This algebra is known as the undeformed $\kappa$-Poincare algebra \cite{hopf}.

The Lorentz generators of the undeformed $\kappa$-Poincare algebra \cite{hopf} satisfy
\begin{equation}\label{CFK7}
 [M_{\mu\nu},M_{\lambda\rho}]=M_{\mu\rho}\eta_{\nu\lambda}-M_{\nu\rho}\eta_{\mu\lambda}-M_{\mu\lambda}\eta_{\nu\rho}+M_{\nu\lambda}\eta_{\mu\rho}.
\end{equation} 
One demands the commutation relation between the Lorentz generators and the $\kappa$-deformed space-time coordinate to be linear in $M_{\mu\nu}$ and $\hat{x}_{\mu}$, i.e.,
\begin{equation}\label{CFK8a}
 [M_{\mu\nu},\hat{x}_{\lambda}]=\hat{x}_{\mu}\eta_{\nu\lambda}-\hat{x}_{\nu}\eta_{\mu\lambda}+ia(M_{0\mu}\eta_{\nu\lambda}-M_{0\nu}\eta_{\mu\lambda}),
\end{equation}
and then using Jacobi identities, one finds the explicit form of the Lorentz generators of the undeformed $\kappa$-Poincare algebra as
\begin{equation}\label{CFK8}
\begin{split}
 M_{ij}=&x_i\partial_j-x_j\partial_i ~,\\
 M_{i0}=&x_i\partial_0\varphi\frac{e^{2A}-1}{2A}-x_0\partial_i\frac{1}{\varphi}+iax_i\partial_k^2\frac{1}{2\varphi}-iax_k\partial_k\partial_i\frac{\gamma}{\varphi}.
\end{split} 
\end{equation}
Note that the usual partial derivative, i.e., $\partial_{\mu}$, does not transform as a $4$-vector under the undeformed $\kappa$-Poincare algebra. This necessitates the introduction of a new  derivative referred to as Dirac derivative, $D_{\mu}$ \cite{hopf}, which transforms as a $4$-vector under this algebra, that is,
\begin{equation}\label{CFK9}
\begin{split}  
 [M_{\mu\nu},D_{\lambda}]=&D_{\mu}\eta_{\nu\lambda}-D_{\nu}\eta_{\mu\lambda},\\
 [D_{\mu},D_{\nu}]=&0,
\end{split}
\end{equation}
where the components of the Dirac derivative are given to be
\begin{equation}\label{CFK10}
\begin{split}
 D_0=&\partial_0\frac{\sinh A}{A}+ia\partial_k^2\frac{e^{-A}}{2\varphi^2},\\
 D_i=&\partial_i\frac{e^{-A}}{\varphi},
\end{split}
\end{equation}
satisfying
\begin{equation}\label{CFK10a}
 [D_{\mu},\hat{x}_{\nu}]=\eta_{\mu\nu}(iaD_0+\sqrt{1+a^2D_{\alpha}D^{\alpha}})+ia\eta_{\mu 0}D_{\nu}.
\end{equation}
The quadratic Casimir invariant (Casimir operator) corresponding to the undeformed $\kappa$-Poincare algebra is
\begin{equation}\label{CFK11}
 D_{\mu}D^{\mu}=\Box\left(1+\frac{a^2}{4}\Box\right),
\end{equation} 
where $\Box$ represents the $\kappa$-deformed Laplacian,
\begin{equation}\label{CFK12}
 \Box=\partial_k^2\frac{e^{-A}}{2\varphi^2}-\partial_0^2\frac{2(1-\cosh A)}{A^2}.
\end{equation}
 $D_\mu$ and $M_{\mu\nu}$ thus define undeformed $\kappa$-Poincare algebra and the corresponding quadratic Casimir invariant, $D_\mu D^\mu=0$ in the momentum space, which is nothing but the deformed dispersion relation, is
\begin{equation}
\frac{4}{a^2}\sinh^2(\frac{A}{2}) -p_ip_i \frac{e^{-A}}{\varphi^2(A)}-m^2c^2 +\frac{a^2}{4}\left[\frac{4}{a^2}\sinh^2(\frac{A}{2}) -p_ip_i \frac{e^{-A}}{\varphi^2(A)}\right]^2=0,\label{dispersion}
\end{equation}
where $p_i$ are the momentum components corresponding to the commutative coordinates.

\section{Deformed metric in $\kappa$-space-time}
In this Section, we construct the Rindler metric\footnote{An effective foundation for comprehending how acceleration affects space-time is provided by Rindler space-time. The main focus of special relativity is on inertial frames or frames that move at a constant speed. Rindler space-time enables us to investigate how constant acceleration affects perceptions of motion, space, and time.} \cite{rindler} in the $\kappa$-deformed space-time using the generalised commutation relation between the $\kappa$-deformed phase-space coordinates \cite{zuhair1}. The coordinates of $\kappa$-deformed space-time satisfy the commutation relations given in Eq.(\ref{intro2}). The generalised commutation relation for the $\kappa$-deformed phase space coordinates is \cite{hopf,34,zuhair1,kappa-geod},
\begin{equation}\label{N1}
 [\hat{x}_{\mu},\hat{P}_{\nu}]=i\hat{g}_{\mu\nu}(\hat{x}),
\end{equation} 
where $\hat{g}_{\mu\nu}$ is the $\kappa$-deformed metric and it is a function of the $\kappa$-deformed space-time coordinate $\hat{x}_{\mu}$. 
We choose a specific realisation for the $\kappa$-deformed phase-space coordinates as \cite{zuhair1},
\begin{equation}\label{N2}
 \hat{x}_{\mu}=x_{\alpha}\varphi^{\alpha}_{\mu}, \,\hat{P}_{\mu}=g_{\alpha\beta}(\hat{y})p^{\beta}\varphi^{\alpha}_{\mu},
\end{equation}
where $\hat{P}_{\mu}$ is the $\kappa$-deformed generalised momenta and $p_{\mu}$ is the canonical conjugate momenta corresponding to the commutative coordinate $x_{\mu}$. In the commutative limit, i.e., $a\to 0$, we obtain $\hat{x}_{\mu}\to x_{\mu}$ and $\hat{P}_{\mu}\to p_{\mu}$. Substituting Eq.(\ref{N2}) in the $\kappa$-deformed space-time commutation relations, i.e., $[\hat{x}_0,\hat{x}_i]=ia\hat{x}_i,~[\hat{x}_i,\hat{x}_j]=0$, we find $\varphi_{\mu}^{\alpha}$ as
\begin{equation}\label{N3}
 \varphi _0^0=1, \, \varphi _i^0=0, \, \varphi_0^i=0, \, \varphi _j^i=\delta _j^i e^{-\frac{ap^0}{2}}. 
\end{equation}
Note that we have introduced another set of $\kappa$-deformed space-time coordinates $\hat{y}_{\mu}$ in Eq.(\ref{N2}). The coordinates $\hat{y}_{\mu}$ are also assumed to satisfy the $\kappa$-deformed space-time commutation relation as $[\hat{y}_0,\hat{y}_i]=ia\hat{y}_i,~[\hat{y}_i,\hat{y}_j]=0$. This $\hat{y}_{\mu}$ is assumed to commute with $\hat{x}_{\mu}$, i.e., $[\hat{y}_{\mu},\hat{x}_{\nu}]=0$. These new coordinates are introduced only for calculational simplification \cite{zuhair1}. The $g_{\alpha\beta}(\hat{y})$ appearing in Eq.(\ref{N2}) has the same functional form as the metric in the commutative coordinate, but $x_{\mu}$ replaced with non-commutative coordinate $\hat{y}_{\mu}$. We now express $\hat{y}_{\mu}$ in terms of the commutative coordinate and its conjugate momenta as
\begin{equation}\label{N4}
 \hat{y}_{\mu}=x_{\alpha}\phi_{\mu}^{\alpha}.
 \end{equation}
Using $[\hat{y}_0,\hat{y}_i]=ia\hat{y}_i,~[\hat{y}_i,\hat{y}_j]=0$ and $[\hat{x}_{\mu},\hat{y}_{\nu}]=0$, one obtains $\phi_{\mu}^{\alpha}$ as  \cite{zuhair1}

\begin{equation}\label{N5}
 \phi_{0}^{0}=1,~\phi_{i}^{0}=0,~\phi_{0}^{i}=-\frac{a}{2}p^i,~\phi_{i}^{j}=\delta_{i}^{j}.
\end{equation}
Thus, the explicit form of $\hat{y}_{\mu}$ are
\begin{equation}\label{N6}
 \hat{y}_0=x_0-\frac{a}{2}x_jp^j,~~
\hat{y}_i=x_i.
 \end{equation}
Using the above in Eq.(\ref{N2}) and substituting $\hat{x}_{\mu}$ and $\hat{P}_{\mu}$ in Eq.(\ref{N1}), the $\kappa$-deformed metric is \cite{zuhair1}
\begin{equation}\label{N7}
 [\hat{x}_{\mu},\hat{P}_{\nu}] \equiv i\hat{g}_{\mu\nu}=ig_{\alpha\beta}(\hat{y})\Big(p^{\beta}\frac{\partial \varphi^{\alpha}_{\nu}}{\partial p^{\sigma}}\varphi_{\mu}^{\sigma}+\varphi_{\mu}^{\alpha}\varphi_{\nu}^{\beta}\Big). \end{equation}
Note that $g_{\mu\nu}(\hat{y})$ can be obtained by replacing the commutative coordinates with the $\kappa$-deformed coordinates for any given commutative metric.

Substituting Eq.(\ref{N3}) in Eq.(\ref{N7}) we find the explicit form of the components of $\hat{g}_{\mu\nu}$ as
\begin{equation}\label{N9}
\begin{aligned}
\hat{g}_{00}&=g_{00}(\hat{y}),\\
\hat{g}_{0i}&=g_{0i}(\hat{y})\big(1-ap^0\big)-\frac{a}{2}g_{im}(\hat{y})p^m,\\ 
\hat{g}_{i0}&=g_{i0}(\hat{y})\big(1-\frac{a}{2}p^0\big),\\
\hat{g}_{ij}&=g_{ij}(\hat{y})\big(1-ap^0\big).
\end{aligned}
\end{equation}
Note here that we consider terms valid up to the first order in $a$ only. From Eq.(\ref{N6}) we see that $g_{\mu\nu}(\hat{y}_i)=g_{\mu\nu}(x_i)$. 

In this study, we investigate a falling Casimir apparatus in the presence of a uniform gravitational field in the background of $\kappa$-space-time. Thus, we take Rindler space-time \cite{rindler} as the background, where $ x\equiv ~ (\tau,x,y,\xi)$ represents the coordinates. The proper acceleration of the particle in Rindler space-time is given by $\frac{1}{\xi}$. Background metric for the Rindler space-time is $g^{(0)}_{\mu\nu}=\text{diag}(-\xi^{2},+1,+1,+1)$. Using this expression in Eq.(\ref{N9}) we find the $\kappa$-deformed metric in the Rindler space-time as

\bea
\hat{g}_{\mu\nu} &=& g^{(0)}_{\mu\nu}+g^{(1)}_{\mu\nu}(a) \nonumber \\ 
&=&\left[\begin{array}{cccc}
-\xi^{2} & 0 & 0 & 0	\\
0 & 1 & 0 & 0	\\
0 & 0 & 1 & 0	\\
0 & 0 & 0 & 1
\end{array}\right] + a\left[\begin{array}{cccc}
0 & -\frac{p^{1}}{2} & -\frac{p^{2}}{2} & -\frac{p^{3}}{2}	\\
0 & -p^{0} & 0 & 0	\\
0 & 0 & -p^{0} & 0	\\
0 & 0 & 0 & -p^{0}
\end{array}\right]. \label{N10}
\eea
Note that the $\hat{g}_{00}$ component does not get $a$-dependent corrections while $\hat{g}_{ii}$ get modified, so do $\hat{g}_{0i}$ components. In the above equation, $p^{0},p^{1},p^{2}$ and $p^{3}$ are energy and momentum scales related to the probe of the non-commutative space-time. The line element in the Rindler-space-time is $ds^{2}=-\xi^{2}d\tau^{2}+dx^{2}+dy^{2}+d\xi^{2}$ and expression of the momentum components are $p^{i}=m\frac{dx^{i}}{d\tau}$, where $i=1,2,3$. Here $d\tau$ is dimensionless and $m$ has an inverse dimension of $dx^{i}$. Thus, $p^{1}$ is dimensionless. $\frac{p^{1}}{\xi}=\frac{m}{\xi}\frac{dx}{d\tau}$ has dimension of $\text{Length}^{-1}$ and $a\frac{p^{1}}{\xi}$ is dimensionless. Note here that we have considered corrections terms to the metric valid only up to the first nonvanishing order in deformation parameter $a$. In the limit $a \rightarrow 0$, we get back the commutative form of the Rindler metric. We also observe that the $\kappa$-deformed Rindler metric is not symmetric because of the presence of off-diagonal correction terms.

Using Eq.(\ref{CFK6a}) we find the coordinates for $\kappa$-deformed rindler space-time as $\hat{t}=\hat{\xi}\sinh \hat{\tau}\Rightarrow t=(1-ap^{0})\xi \sinh \tau$ and $\hat{z}=\hat{\xi}\cosh\hat{\tau}\Rightarrow z=\xi \cosh \tau$, valid upto first order in deformation parameter $a$. These coordinates satisfy equation of $\kappa$-deformed hyperbola, that is,
\be
z^{2}- (1+ap^{0})t^{2}=\xi^{2}.  \label{khyp} 
\ee
Thus, the uniformly accelerated observer/particle follows deformed hyperbolic paths where the time coordinate alone gets $a$-dependent correction. This correction is also dependent on the energy $p^{0}$ of the observer/particle.

\section{Lagrangian, Equation of Motion and Energy-Momentum tensor}
In this Section, we construct the Lagrangian describing the scalar field in the presence of the Casimir apparatus in $\kappa$-deformed Rindler space-time. Using this Lagrangian, we find the equation of motion for the scalar field in the presence of Casimir plates and the corresponding $\kappa$-deformed energy-momentum tensor. Using the expression of the quadratic Casimir invariant of the undeformed $\kappa$-Poincare algebra given in Eq.(\ref{CFK11}) we find the generalized massless Klein-Gordon equation in the $\kappa$-deformed space-time to be
\be
 \Box\left(1+\frac{a^2}{4}\Box\right)\phi=0, \label{L1}
\ee
where $\Box$ is given in Eq.(\ref{CFK12}). From the above equation, we find (for the realization $\varphi=e^{-\frac{A}{2}}$) equation of motion valid up to the first-nonvanishing parameter in deformation parameter $a$ as
\be
 \Big(\partial_{\mu}\partial^{\mu}(1+\frac{a^{2}}{4}\partial_{\alpha}\partial^{\alpha})+\frac{a^{2}}{12}\partial_{0}^{4}\Big)\phi=0. \label{L2}
\ee
The above equation of motion can be obtained from the following Lagrangian
\be
L=-\frac{1}{2}\partial_{\mu}\phi\partial^{\mu}\phi+\frac{a^{2}}{8}\partial_{\mu}\partial^{\mu}\phi\partial_{\alpha}\partial^{\alpha}\phi+\frac{a^{2}}{24}\partial_{0}^{2}\phi\partial_{0}^{2}\phi,  \label{L3}
\ee 
which we can re-express as
\be
L= -\frac{1}{2}\eta^{\mu\nu}\partial_{\mu}\phi\partial_{\nu}\phi+\frac{a^{2}}{8}\eta^{\mu\nu}\partial_{\mu}\partial_{\nu}\phi\eta^{\alpha\beta}\partial_{\alpha}\partial_{\beta}\phi+\frac{a^{2}}{24}\delta_{\mu 0}\delta_{\nu 0}\partial^{\mu}\partial^{\nu}\phi \delta_{\alpha 0}\delta_{\beta 0}\partial^{\alpha}\partial^{\beta}\phi. \label{L4}
\ee
Replacing $\eta^{\mu\nu}$ in the above equation with $\hat{g}^{\mu\nu}$given in Eq.(\ref{N10}) we find the Lagrangian for scalar theory valid up to the first order in deformation parameter $a$ in the $\kappa$-deformed Rindler space-time as
\bea
 \mathcal{L}^{0} &=&-\frac{1}{2}g^{(0){\mu\nu}}\partial_{\mu}\phi \partial_{\nu}\phi-\frac{1}{2}g^{(1){\mu\nu}}(a)\partial_{\mu}\phi \partial_{\nu}\phi \nonumber \\
 &=& \frac{1}{2\xi^{2}}(\partial_{0}\phi)^{2}-\frac{1}{2}\Big((\partial_{x}\phi)^{2}+(\partial_{y}\phi)^{2}+(\partial_{\xi}\phi)^{2}\Big)\nonumber \\&&-\frac{ap^{0}}{2}\Big((\partial_{x}\phi)^{2}+(\partial_{y}\phi)^{2}+(\partial_{\xi}\phi)^{2}\Big)+\frac{ap^{1}}{4\xi^{2}}\partial_{0}\phi\partial_{x}\phi + \frac{ap^{2}}{4\xi^{2}}\partial_{0}\phi\partial_{y}\phi + \frac{ap^{3}}{4\xi^{2}}\partial_{0}\phi\partial_{\xi}\phi. \label{L5}
\eea
To study the Casimir effect for a single plate and for two parallel plates, one needs to introduce these plates through their interactions with fields. The interactions of the plates for both cases (single plate and parallel plates) are described by the interaction Lagrangian given by
\be
\mathcal{L}_{int}=-\frac{1}{2}V(x)\phi^{2}  \label{L6}
\ee
where 
\bea
V(x)&=&\lambda_{1}\delta(\xi-\xi_{1}), ~~~~~~~~~~~~~~~~~~~~~~~~ \text{for~single~plate}, \nonumber\\
    &=&\lambda_{1}\delta(\xi-\xi_{1})+\lambda_{2}\delta(\xi-\xi_{2}), ~~~~~~\text{for~parallel~plates}. \label{L7}
\eea
Here $\lambda_{1}$ and $\lambda_{2}$ are coupling constants with dimension of $\text{length}^{-1}$.
Thus, using Eq.(\ref{L5}) and Eq.(\ref{L6}) we find complete Lagrangian in the $\kappa$-deformed Rindler space-time as
\be
\mathcal{L}=\mathcal{L}^{0}+\mathcal{L}_{\text{int}}  \label{L8}
\ee
and we also find the corresponding action  in $\kappa$-deformed Rindler space-time as
\be
 \mathcal{S}_{\text{$\kappa$-Rindler}}=\int d^{4}x \sqrt{-\hat{g}}\mathcal{L}(\phi,\partial_{\mu}\phi,a), \label{L9}
\ee
where $\hat{g}=\text{det}~\hat{g}_{\mu\nu}=-\xi^{2}(1-3ap^{0})$ valid upto first order in $a$. In the limit $a \rightarrow 0$, we get back the action for commutative case \cite{shajesh2a,shajesh2b}. Next, by varying the above action, we find the equation of motion as
\be
\left\lbrace-\frac{1}{\xi^2}\partial_{\tau}^{2}+\frac{1}{\xi}\partial_{\xi}(\xi\partial_{\xi})+\nabla^{2}_{\perp}-V(x)+ ap^{0}\Big(\frac{1}{\xi}\partial_{\xi}(\xi\partial_{\xi})+\nabla^{2}_{\perp}\Big)-a\Big(\frac{p^{1}}{2\xi^2}\partial_{\tau}\partial_{x}+\frac{p^{2}}{2\xi^2}\partial_{\tau}\partial_{y}+\frac{p^{3}}{2\xi^{2}}\partial_{\tau}\partial_{\xi}\Big)\right\rbrace \phi=0, \label{L10}
\ee
%($\clubsuit$  Thus in the above equation we find $\frac{1}{\xi}\partial_{x}$ has dimention of $Length^{-2}$.)
where $\nabla^{2}_{\perp}=\partial_{x}^{2}+\partial_{y}^{2}$ and explicit form of $V(x)$ is given in Eq.(\ref{L7}). Note that we have taken correction terms valid up to the first order in deformation parameter $a$. By varying the action given in Eq.(\ref{L9}) with respect of the background metric given in Eq.(\ref{N10}) we find the expression for the energy-momentum tensor as
\be
\hat{T}_{\mu\nu}=2\Big(-\frac{\partial \mathcal{L}}{\partial \hat{g}^{\mu\nu}(a)}+\frac{1}{2}\hat{g}_{\mu\nu}(a)\mathcal{L}\Big), \label{L11}
\ee
Here, we observe that the presence of off-diagonal terms in the $\kappa$-deformed Rindler metric in Eq.(\ref{N10}) makes the above expression of energy-momentum asymmetric. Therefore, we explicitly symmetrize the energy-momentum tensor and obtain
\be
 \hat{T}_{\mu\nu}=\frac{1}{2}\left\lbrace\Big(-\frac{\partial \mathcal{L}}{\partial \hat{g}^{\mu\nu}(a)}+\frac{1}{2}\hat{g}_{\mu\nu}(a)\mathcal{L}\Big)+\Big(-\frac{\partial \mathcal{L}}{\partial \hat{g}^{\nu\mu}(a)}+\frac{1}{2}\hat{g}_{\nu\mu}(a)\mathcal{L}\Big)\right\rbrace. \label{L12}
\ee
Next, using Eq.(\ref{L5}), Eq.(\ref{L6}), and Eq.(\ref{L8}) in the above equation, we find the expression for the energy-momentum tensor to be
\be
\hat{T}_{\mu\nu}=\partial_{\mu}\phi\partial_{\nu}\phi+\frac{1}{2}\Big(\hat{g}_{\mu\nu}(a)+\hat{g}_{\nu\mu}(a)\Big)\mathcal{L}, \label{L13} 
\ee
where $\mathcal{L}$ is given in Eq.(\ref{L8}). We have taken correction terms that are valid only up to the first order in $a$. In the commutative limit, i.e., $a \rightarrow 0$, we get back the commutative expression of energy-momentum tensor \cite{shajesh2a,shajesh2b}.

\section{Green's function solutions to the equations of motions}
In this Section, we solve the equation of motion given in Eq.(\ref{L10}) for both single plate and two parallel plates and find the Green's function in different regions. We also find the Minkowski limit (weak field approximation) of those solutions, which are required in the next Section. Since the force acting on the Casimir apparatus in the presence of the gravitational field will be calculated using Green's function, we obtain Green's function by solving the equation of motion given in Eq.(\ref{L10}) for the following two cases.

\subsection{Deformed Green's function for Single plate}
For single-plate situated at $\xi=\xi_{1}$, using Eq.(\ref{L7}) in Eq.(\ref{L10}), we find the Euler-Lagrange equation and the corresponding Green's function satisfies
 %\be
 %
%\bigg\{-\frac{1}{\xi^2}\partial_{\tau}^{2}+\frac{1}{\xi}\partial_{\xi}(\xi\partial_{\xi})+\nabla^{2}_{\perp}-\lambda_{1}\delta(\xi-\xi_{1})+ ap^{0}\Big(\frac{1}{\xi}\partial_{\xi}(\xi\partial_{\xi})+\nabla^{2}_{\perp}\Big)-a\Big(\frac{p^{1}}{2\xi^2}\partial_{\tau}\partial_{x}+\frac{p^{2}}{2\xi^2}\partial_{\tau}\partial_{y}+\frac{p^{3}}{2\xi^{2}}\partial_{\tau}\partial_{\xi}\Big)\bigg\} \phi=0, \label{L14}
%\ee
\begin{multline}
-\bigg\{-\frac{1}{\xi^2}\partial_{\tau}^{2}+\frac{1}{\xi}\partial_{\xi}(\xi\partial_{\xi})+\nabla^{2}_{\perp}-\lambda_{1}\delta(\xi-\xi_{1})+ ap^{0}\Big(\frac{1}{\xi}\partial_{\xi}(\xi\partial_{\xi})+\nabla^{2}_{\perp}\Big)\\-a\Big(\frac{p^{1}}{2\xi^2}\partial_{\tau}\partial_{x}+\frac{p^{2}}{2\xi^2}\partial_{\tau}\partial_{y}+\frac{p^{3}}{2\xi^{2}}\partial_{\tau}\partial_{\xi}\Big)\bigg\} \tilde{G}(x,x^{\prime},a)=(1+\frac{3}{2}ap^{0})\frac{\delta(\tau-\tau^{\prime})\delta(\xi-\xi^{\prime})\delta(x_{\perp}-x_{\perp}^{\prime})}{\xi}. \label{L15}
\end{multline}
Note that correction terms on the right-hand side are coming from $\sqrt{-g}=(1-\frac{3}{2}ap^{0})\xi$.
Since in our case $V(x)=\lambda_{1}\delta(\xi-\xi_{1})$ has only $\xi$ dependence, we can work with reduced Green's function $\tilde{g} (\xi,\xi^{\prime},a)$. Thus, the Fourier transformation of Green's function is expressed as
\be
\tilde{G}(x,x^{\prime},a)=\int_{\infty}^{\infty} \frac{d\omega d^{2}k_{\perp}}{8\pi^{3}}e^{-i\omega(\tau-\tau^{\prime})}e^{i{\bf k_{\perp}}\cdot({\bf x_{\perp}}-{\bf x_{\perp}^{\prime}})}\tilde{g}(\xi,\xi^{\prime},a). \label{L16}
\ee
Using Eq.(\ref{L16}) in Eq.(\ref{L15}) we find
\begin{multline}
-\bigg\{\frac{1}{\xi}\partial_{\xi}(\xi\partial_{\xi})+\frac{\omega^{2}}{\xi^{2}}-k_{\perp}^{2}-\lambda_{1}\delta(\xi-\xi_{1})+ ap^{0}\Big(\frac{1}{\xi}\partial_{\xi}(\xi\partial_{\xi})-k_{\perp}^{2}\Big)\\-a\Big(\frac{p^{1}}{2\xi^2}(\omega k_{x})+\frac{p^{2}}{2\xi^2}(\omega k_{y})+\frac{p^{3}}{2\xi^{2}}(-i \omega)\partial_{\xi}\Big)\bigg\} \tilde{g}(\xi,\xi^{\prime},a)=(1+\frac{3}{2}ap^{0})\frac{\delta(\xi-\xi^{\prime})}{\xi}. \label{L17}
\end{multline}
We solve this equation  perturbatively assuming the solution $\tilde{g}(\xi,\xi^{\prime},a)$ to be of the form
\be
\tilde{g}(\xi,\xi^{\prime},a)=\tilde{g}^{(0)}(\xi,\xi^{\prime})+ap^{0}\tilde{g}^{(1)}(\xi,\xi^{\prime}). \label{L18}
\ee
Substituting Eq.(\ref{L18}) in Eq.(\ref{L17}), we obtain the following equations
\be
 -\bigg\{\frac{1}{\xi}\partial_{\xi}(\xi\partial_{\xi})+\frac{\omega^{2}}{\xi^{2}}-k_{\perp}^{2}-\lambda_{1}\delta(\xi-\xi_{1})\bigg\}\tilde{g}^{(0)}(\xi,\xi^{\prime})=\frac{\delta(\xi-\xi^{\prime})}{\xi}, \label{L19}
\ee
and
\begin{multline}
 -\bigg[\Big(\frac{1}{\xi}\partial_{\xi}(\xi\partial_{\xi})+\frac{\omega^{2}}{\xi^{2}}-k_{\perp}^{2}-\lambda_{1}\delta(\xi-\xi_{1})\Big)\tilde{g}^{(1)}(\xi,\xi^{\prime}) \\ +\Big\{\ \frac{1}{\xi}\partial_{\xi}(\xi\partial_{\xi})-k_{\perp}^{2} -\frac{1}{p^{0}}\Big(\frac{p^{1}}{2\xi^2}(\omega k_{x})+\frac{p^{2}}{2\xi^2}(\omega k_{y})+\frac{p^{3}}{2\xi^{2}}(-i \omega)\partial_{\xi}\Big)\Big\}\ \tilde{g}^{(0)}(\xi,\xi^{\prime})\bigg]=\frac{3}{2}\frac{\delta(\xi-\xi^{\prime})}{\xi}. \label{L20}
\end{multline}
We solve Eq.(\ref{L19}) after replacing $\omega=i\zeta$ and find the solution for  $\tilde{g}^{(0)}(\xi,\xi^{\prime})$ in different regions for a single plate placed at $\xi=\xi_1$ as \cite{shajesh2a,shajesh2b}
\bea 
\tilde{g}^{(0)}(\xi,\xi^{\prime}) &=& I_{\zeta}(k_{\perp}\xi_{<})K_{\zeta}(k_{\perp }\xi_{>}) - \frac{\lambda_{1}\xi_{1}K_{\zeta}^{2}(k_{\perp}\xi_{1})I_{\zeta}(k_{\perp} \xi)I_{\zeta}(k_{\perp} \xi^{\prime})}{1+\lambda_{1}\xi_{1}I_{\zeta}(k_{\perp} \xi_{1})K_{\zeta}(k_{\perp} \xi_{1})}; ~~~~\text{where}~ \left\lbrace\xi,\xi^{\prime}\right\rbrace < \xi_{1},\label{L21}\\
&=& I_{\zeta}(k_{\perp}\xi_{<})K_{\zeta}(k_{\perp }\xi_{>}) - \frac{\lambda_{1}\xi_{1}I_{\zeta}^{2}(k_{\perp}\xi_{1})K_{\zeta}(k_{\perp} \xi)K_{\zeta}(k_{\perp} \xi^{\prime})}{1+\lambda_{1}\xi_{1}I_{\zeta}(k_{\perp} \xi_{1})K_{\zeta}(k_{\perp} \xi_{1})}; ~~~\text{where}~ \left\lbrace\xi,\xi^{\prime}\right\rbrace > \xi_{1}.\label{L22}
\eea
Here, $I$ and $K$ are modified Bessel functions of the first kind and second kind, respectively.
Using Eq.(\ref{L21}) and Eq.(\ref{L22}) in Eq.(\ref{L20}) and find expression of  $\tilde{g}^{(1)}(\xi,\xi^{\prime})$ as
\begin{multline}
\tilde{g}^{(1)}(\xi,\xi^{\prime})=\frac{1}{2}\tilde{g}^{(0)}(\xi,\xi^{\prime})+\int d\bar{\xi}~\bar{\xi}~\tilde{g}^{(0)}(\xi,\bar{\xi})\bigg\{\frac{\zeta^{2}}{\bar{\xi}^{2}}+\lambda_{1} \delta(\bar{\xi}-\xi_{1})\\-\frac{1}{p^{0}}\Big(\frac{p^{1}}{2\bar{\xi}^2}(\omega k_{x})+\frac{p^{2}}{2\bar{\xi}^2}(\omega k_{y})+\frac{p^{3}}{2\bar{\xi}^{2}}(-i \omega)\partial_{\bar{\xi}}\Big)\bigg\}\tilde{g}^{(0)}(\bar{\xi},\xi^{\prime}). \label{L23}
\end{multline}
$\tilde{g}^{(0)}(\xi,\bar{\xi})$ and $\tilde{g}^{(0)}(\bar{\xi},\xi^{\prime})$ appearing in the second term on the right-hand side of the above equation are also obtained from Eq.(\ref{L21}) and Eq.(\ref{L22}). Note $\bar{\xi}$ takes value between $\xi$ and $\xi^{\prime}$. Integration operation in the second term of the above equation ranges from $\xi$ to $\xi^{\prime}$ when $\xi^{\prime} > \xi$ and it ranges $\xi^{\prime}$ to $\xi$ when $\xi > \xi^{\prime}$. Thus, we observe that delta function $\big(\delta(\bar{\xi}-\xi_{1})\big)$ dependent term inside the integration do not contribute.

Thus, using Eq.(\ref{L21}), Eq.(\ref{L22}), and Eq.(\ref{L23}) in Eq.(\ref{L18}) (with $\omega = i \zeta$) we find the reduced Green's function, for Euler-Lagrange equation given in Eq.(\ref{L17}) valid upto first order in $a$ as
\bea
\tilde{g}(\xi,\xi^{\prime},a)&=&(1+\frac{ap^{0}}{2})\tilde{g}^{(0)}(\xi,\xi^{\prime})+ap^{0}\Bigg[\int d\bar{\xi}~ \bar{\xi} \tilde{g}^{(0)}(\xi,\bar{\xi})\bigg\{\frac{\zeta^{2}}{\bar{\xi}^{2}}+\lambda_{1} \delta(\bar{\xi}-\xi_{1})\nonumber\\&&-\frac{1}{p^{0}}\Big(\frac{p^{1}}{2\bar{\xi}^2}(i \zeta k_{x})+\frac{p^{2}}{2\bar{\xi}^2}(i \zeta k_{y})+\frac{p^{3}}{2\bar{\xi}^{2}}(\zeta)\partial_{\bar{\xi}}\Big)\bigg\}\tilde{g}^{(0)}(\bar{\xi},\xi^{\prime})\Bigg]. \label{L24}
\eea
 We observe here that in finding 
$\tilde{g}(\xi,\xi,a)$ from above expression, only first term will contribute. In the next Section, we will see that the gravitational force acting on the plates can be calculated using $\tilde{g}(\xi,\xi)$. Also, we will see that one needs to take the Minkowski-space limit (weak field approximation) to calculate the force acting on the plates.

 To find the Minkowski-space limit for the reduced Green's function (for details see \cite{shajesh2a,shajesh2b}) we take the limit
 \be
 \xi \rightarrow \infty, ~~ \zeta \rightarrow \infty ~~, \label{L25}
 \ee
such that the ratio $\frac{\zeta}{\xi}$ and differences in $\xi$ ( i.e., $\xi-\xi_{0}$, $\xi^{\prime}-\xi_{0}$ and $\xi_{1}-\xi_{0}$) remain finite. Thus, we find the uniform asymptotic expansion (Debye expansion) for the modified Bessel function as
\be
 I_{\zeta}(k_{\perp}\xi)~\sim ~\sqrt{\frac{t}{2 \pi \zeta}} e^{\zeta \eta(\xi)} \sum_{n=0}^{\infty} \frac{1}{\zeta^{n}}u_{n}(t) ~~~\text{and}~~~K_{\zeta}(k_{\perp}\xi)~\sim ~\sqrt{\frac{\pi t}{2 \zeta}} e^{-\zeta \eta(\xi)} \sum_{n=0}^{\infty} \frac{(-1)^{n}}{\zeta^{n}}u_{n}(t), \label{L26}
\ee
where 
\be
t=\frac{1}{\sqrt{1+\frac{k_{\perp}^{2}\xi^{2}}{\zeta^{2}}}} ~~~\text{and}~~~ \eta(\xi)=\sqrt{1+\frac{k_{\perp}^{2}\xi^{2}}{\zeta^{2}}} + \text{ln} \Bigg[\frac{\frac{k_{\perp}\xi}{\zeta}}{1+\sqrt{1+\frac{k_{\perp}^{2}\xi^{2}}{\zeta^{2}}}} \Bigg]. \label{L27}
\ee
Here $u_{n}(t)$ are polynomials in t \cite{stegun}. Next we expand the above expressions around arbitrary point $\xi_{0}$ by considering  $\xi-\xi_{0}$, $\xi^{\prime}-\xi_{0}$ and $\xi_{1}-\xi_{0}$ to be finite. Thus, we find
\be
\sqrt{\xi \xi^{\prime}}I_{\zeta}(k_{\perp}\xi)K_{\zeta}(k_{\perp}\xi^{\prime}) \sim \frac{1}{2\bar{k}}e^{\bar{k}(\xi-\xi^{\prime})},~~\text{where}~~\bar{k}^{2}=k_{\perp}^{2}+\frac{\zeta^{2}}{\xi_{0}^{2}} ~~\text{and}~~ \frac{1}{\sqrt{\xi \xi^{\prime}}} \sim \frac{1}{\xi_{0}}. \label{L28}
\ee
%One can write complete form of the above expression in thesis%
As $\xi \rightarrow \infty$,  in the above expression we have kept terms up to first order in $\frac{1}{\xi_{0}}$ only. Next using Eq.(\ref{L28}) in Eq.({\ref{L21}}) and Eq.(\ref{L22}) we find \cite{shajesh2a,shajesh2b}
\bea
 \tilde{g}^{(0)}(\xi,\xi^{\prime}) &\rightarrow &  \frac{g^{(0)}(\xi,\xi^{\prime})}{\xi_{0}}\nonumber \\ &=& \frac{1}{\xi_{0}}\Big\{\frac{1}{2\bar{k}}e^{-\bar{k}|\xi-\xi^{\prime}|}-\frac{\lambda_{1}}{\lambda_{1}+2\bar{k}}\frac{1}{2 \bar{k}}e^{\bar{k}(\xi-\xi_{1})}e^{-\bar{k}(\xi^{\prime}-\xi_{1})}\Big\},~~\text{where}~\left\lbrace\xi,\xi^{\prime}\right\rbrace < \xi_{1}, \nonumber \\
 &=&\frac{1}{\xi_{0}}\Big\{\frac{1}{2\bar{k}}e^{-\bar{k}|\xi-\xi^{\prime}|}-\frac{\lambda_{1}}{\lambda_{1}+2\bar{k}}\frac{1}{2 \bar{k}}e^{-\bar{k}(\xi-\xi_{1})}e^{\bar{k}(\xi^{\prime}-\xi_{1})}\Big\},~~\text{where}~\left\lbrace\xi,\xi^{\prime}\right\rbrace >\xi_{1}. \label{L29}
\eea
Thus, using the above expressions in Eq.(\ref{L24}) reduced Green's function ($g(\xi,\xi^{\prime}, a)$) is obtained in the Minkowski-space limit.

\subsection{Deformed Green's function for two parallel plates}
Next, we consider two parallel plates placed at $\xi=\xi_{1}$ and $\xi=\xi_{2}$, falling rigidly \cite{born} in the presence of the gravitational field in such a way that the distance between those two plates remains same. Thus, for two parallel plates, we use Eq.(\ref{L7}) in Eq.(\ref{L10}) and find that the corresponding Green's function obey
%the Euler-Lagrangian equation and
%\begin{equation}
%\end{equation}
 %\begin{multline}
%\bigg\{-\frac{1}{\xi^2}\partial_{\tau}^{2}+\frac{1}{\xi}\partial_{\xi}(\xi\partial_{\xi})+\nabla^{2}_{\perp}-\lambda_{1}\delta(\xi-\xi_{1})-\lambda_{2}\delta(\xi-\xi_{2})+ ap^{0}\Big(\frac{1}{\xi}\partial_{\xi}(\xi\partial_{\xi})+\nabla^{2}_{\perp}\Big) \\-a\Big(\frac{p^{1}}{2\xi^2}\partial_{\tau}\partial_{x}+\frac{p^{2}}{2\xi^2}\partial_{\tau}\partial_{y}+\frac{p^{3}}{2\xi^{2}}\partial_{\tau}\partial_{\xi}\Big)\bigg\} \phi=0, \label{L30}
%\end{multline}
\begin{multline}
-\bigg\{-\frac{1}{\xi^2}\partial_{\tau}^{2}+\frac{1}{\xi}\partial_{\xi}(\xi\partial_{\xi})+\nabla^{2}_{\perp}-\lambda_{2}\delta(\xi-\xi_{2})-\lambda_{1}\delta(\xi-\xi_{1})+ ap^{0}\Big(\frac{1}{\xi}\partial_{\xi}(\xi\partial_{\xi})+\nabla^{2}_{\perp}\Big)\\-a\Big(\frac{p^{1}}{2\xi^2}\partial_{\tau}\partial_{x}+\frac{p^{2}}{2\xi^2}\partial_{\tau}\partial_{y}+\frac{p^{3}}{2\xi^{2}}\partial_{\tau}\partial_{\xi}\Big)\bigg\} \tilde{G}(x,x^{\prime},a)=(1+\frac{3}{2}ap^{0})\frac{\delta(\tau-\tau^{\prime})\delta(\xi-\xi^{\prime})\delta(x_{\perp}-x_{\perp}^{\prime})}{\xi}. \label{L31}
\end{multline}
 As in the case of a single plate, here also using the Fourier transformation of Green's function given in Eq.(5.3) in Eq.(\ref{L31}), we find that the reduced Green's function satisfies
%\be
%\tilde{G}(x,x^{\prime},a)=\int_{\infty}^{\infty} \frac{d\omega d^{2}k_{\perp}}{8\pi^{3}}e^{-i\omega(\tau-\tau^{\prime}}e^{ik_{\perp}(x_{\perp}-x_{\perp}^{\prime})}\tilde{g}(\xi,\xi^{\prime},a). \label{L32}
%\ee
\begin{multline}
-\bigg\{\frac{1}{\xi}\partial_{\xi}(\xi\partial_{\xi})+\frac{\omega^{2}}{\xi^{2}}-k_{\perp}^{2}-\lambda_{1}\delta(\xi-\xi_{1})-\lambda_{2}\delta(\xi-\xi_{2})+ ap^{0}\Big(\frac{1}{\xi}\partial_{\xi}(\xi\partial_{\xi})-k_{\perp}^{2}\Big)\\-a\Big(\frac{p^{1}}{2\xi^2}(\omega k_{x})+\frac{p^{2}}{2\xi^2}(\omega k_{y})+\frac{p^{3}}{2\xi^{2}}(-i \omega)\partial_{\xi}\Big)\bigg\} \tilde{g}(\xi,\xi^{\prime},a)=(1+\frac{3}{2}ap^{0})\frac{\delta(\xi-\xi^{\prime})}{\xi}. \label{L33}
\end{multline}
By following the same procedure discussed in the previous subsection for the single plate case, we solve the above equation perturbatively (that is, we use Eq.(\ref{L18})). Thus, we find the reduced Green's function valid up to the first order in $a$ to be
\bea
\tilde{g}(\xi,\xi^{\prime},a)&=&(1+ap^{0}\frac{1}{2})\tilde{g}^{(0)}(\xi,\xi^{\prime})+ap^{0}\Bigg[\int d\bar{\xi}~ \bar{\xi} \tilde{g}^{(0)}(\xi,\bar{\xi})\bigg\{\frac{\zeta^{2}}{\bar{\xi}^{2}}+\lambda_{1} \delta(\bar{\xi}-\xi_{1})+\lambda_{2} \delta(\bar{\xi}-\xi_{2})\nonumber\\&&-\frac{1}{p^{0}}\Big(\frac{p^{1}}{2\bar{\xi}^2}(\omega k_{x})+\frac{p^{2}}{2\bar{\xi}^2}(\omega k_{y})+\frac{p^{3}}{2\bar{\xi}^{2}}(-i \omega)\partial_{\bar{\xi}}\Big)\bigg\}\tilde{g}^{(0)}(\bar{\xi},\xi^{\prime})\Bigg], \label{L34}
\eea
where 

\bea
\tilde{g}^{(0)}(\xi,\xi^{\prime}) &=& I_{<}K_{>}-\frac{\lambda_{1}\xi_{1}K_{1}^{2}+\lambda_{2}\xi_{2}K_{2}^{2}-\lambda_{1}\lambda_{2}\xi_{1}\xi_{2}K_{1}K_{2}(K_{2}I_{1}-K_{1}I_{2})}{\tilde{\Delta}} I I_{\prime}; ~~\text{where}~~\left\lbrace\xi,\xi^{\prime}\right\rbrace <\xi_{1}<\xi_{2}, \nonumber \\\label{L35}\\
&=& I_{<}K_{>}-\frac{\lambda_{2}\xi_{2}K_{2}^{2}(1+\lambda_{1}\xi_{1}K_{1}I_{1})}{\tilde{\Delta}}I I_{\prime}-\frac{\lambda \xi_{1}I_{1}^{2}(1+\lambda_{2}\xi_{2}K_{2}I_{2})}{\tilde{\Delta}}K K_{\prime}\nonumber \\
&& +\frac{\lambda_{1}\lambda_{2}\xi_{1}\xi_{2}I_{1}^{2}K_{2}^{2}}{\tilde{\Delta}}(I K_{\prime}+K I_{\prime});~~~~~~~~~~~~~~~~~~~~~~~~~~~~~~~~~~~~~~~\text{where}~~\xi_{1}<\left\lbrace\xi,\xi^{\prime}\right\rbrace <\xi_{2}, \nonumber \\ \label{L36}\\
&=& I_{<}K_{>}-\frac{\lambda_{1}\xi_{1}I_{1}^{2}+\lambda_{2}\xi_{2}I_{2}^{2}+\lambda_{1}\lambda_{2}\xi_{1}\xi_{2}I_{1}I_{2}(I_{2}K_{1}-I_{1}K_{2})}{\tilde{\Delta}} K K_{\prime}; ~~~~\text{where}~~\left\lbrace\xi,\xi^{\prime}\right\rbrace >\xi_{2}>\xi_{1}.\nonumber\\ \label{L37}
\eea
In the above equations
\be
\tilde{\Delta}=(1+\lambda_{1}\xi_{1}K_{1}I_{1})(1+\lambda_{2}\xi_{2}K_{2}I_{2})-\lambda_{1}\lambda_{2}\xi_{1}\xi_{2}I_{1}^{2}K^{2}_{2} \label{L38}
\ee
and we have used $I_{i}=I_{\zeta}(k_{\perp}\xi_{i})$, $K_{i}=K_{\zeta}(k_{\perp}\xi_{i})$; where $i=1,2$, $I=I_{\zeta}(k_{\perp}\xi)$, $I_{\prime}=I_{\zeta}(k_{\perp}\xi^{\prime})$, $K=K_{\zeta}(k_{\perp}\xi)$ and $K_{\prime}=K_{\zeta}(k_{\perp}\xi^{\prime})$. Note here that $\tilde{g}^{(0)}(\xi,\bar{\xi})$ and $\tilde{g}^{(0)}(\bar{\xi},\xi^{\prime})$ in Eq.(\ref{L34}) can be obtained using Eq.(\ref{L35}), Eq.(\ref{L36}), Eq.(\ref{L37}) and Eq.(\ref{L38}). As in the case of single plate, here also limit of the integration in the third term inside the square bracket in the Eq.(\ref{L34}) is either $\xi \rightarrow \xi^{\prime}$ for $\xi^{\prime}>\xi$ or $\xi^{\prime} \rightarrow \xi$ for $\xi > \xi^{\prime}$. Note that here we have taken correction terms valid upto first order in deformation parameter $a$. Thus the limit $a \rightarrow 0$  gives us the reduced Green's function in the commutative case \cite{shajesh2a,shajesh2b}.

Next by following the same procedure discussed in the previous subsection (for details see Eq.(\ref{L25}) to Eq.(\ref{L28})), we take the Minkowski-space limit of the Eq.(\ref{L35}) to Eq.(\ref{L38}) and find
\bea
&\tilde{g}^{(0)}(\xi,\xi^{\prime}) \rightarrow  \frac{g^{(0)}(\xi,\xi^{\prime})}{\xi_{0}}& \nonumber \\
&= \frac{1}{\xi_{0}}\Bigg[\frac{1}{2\bar{k}}e^{-\bar{k}|\xi-\xi^{\prime}|}-\frac{1}{2\bar{k}\Delta}e^{\bar{k}(\xi+\xi^{\prime}-2\xi_{1})}\bigg\{\frac{\lambda_{1}}{2\bar{k}}\Big(1+\frac{\lambda_{2}}{2\bar{k}}\Big)e^{2\bar{k}L} + \frac{\lambda_{2}}{2\bar{k}}\Big(1-\frac{\lambda_{1}}{2\bar{k}}\Big)\bigg\}  \Bigg],~~~\text{where}~\left\lbrace\xi,\xi^{\prime}\right\rbrace <\xi_{1}<\xi_{2}.& \nonumber \\ \label{L39}
& = \frac{1}{\xi_{0}}\Bigg[\frac{1}{2\bar{k}}e^{-\bar{k}|\xi-\xi^{\prime}|}-\frac{1}{2\bar{k}\Delta}\bigg\{\frac{\lambda_{1}}{2\bar{k}}\Big(1+\frac{\lambda_{2}}{2\bar{k}}\Big)e^{-\bar{k}(\xi+\xi^{\prime}-2\xi_{2})} + \frac{\lambda_{2}}{2\bar{k}}\Big(1+\frac{\lambda_{1}}{2\bar{k}}\Big)e^{\bar{k}(\xi+\xi^{\prime}-2\xi_{1})}-\frac{\lambda_{1}\lambda_{2}}{4\bar{k}^{2}}2 \cosh \bar{k}(\xi-\xi^{\prime})\bigg\}  \Bigg],& \nonumber \\&~~~~~~~~~~~~~~~~~~~~~~~~~~~~~~~~~~~~~~~~~~~~~~~~~~~~~~~~~~~~~~~~~~~~~~~~~~~~~~~~~~~~~~~~~~~~~~\text{where}~\xi_{1}<\left\lbrace\xi,\xi^{\prime}\right\rbrace <\xi_{2}. &\nonumber \\ \label{L40}
&= \frac{1}{\xi_{0}}\Bigg[\frac{1}{2\bar{k}}e^{-\bar{k}|\xi-\xi^{\prime}|}-\frac{1}{2\bar{k}\Delta}e^{-\bar{k}(\xi+\xi^{\prime}-2\xi_{2})}\bigg\{\frac{\lambda_{1}}{2\bar{k}}\Big(1-\frac{\lambda_{2}}{2\bar{k}}\Big) + \frac{\lambda_{2}}{2\bar{k}}\Big(1+\frac{\lambda_{1}}{2\bar{k}}\Big)e^{2\bar{k}L}\bigg\}  \Bigg],~\text{where}~\left\lbrace\xi,\xi^{\prime}\right\rbrace >\xi_{2}>\xi_{1}.& \nonumber \\ \label{L41}
\eea
In the above expressions $L=\xi_{2}-\xi_{1}$ and
\be
\Delta=\Big(1+\frac{\lambda_{a}}{2\bar{k}}\Big)\Big(1+\frac{\lambda_{2}}{2\bar{k}}\Big)e^{2\bar{k}L}-\frac{\lambda_{1} \lambda_{2}}{4\bar{k}^{2}}. \label{L42} 
\ee
These expressions for $g^{(0)}(\xi,\xi^{\prime})$ (reduced Green's function for commutative case in Minkowski-space-time) exactly match with the result given in \cite{shajesh2a,shajesh2b}. Thus, in the Minkowski-space limit, using the above equations, (that is, Eq.(\ref{L41}) and Eq.(\ref{L42})) in Eq.(\ref{L34}) we obtain the expressions of the reduced Green's function in different regions.

\section{Force acting on the plates}
In this Section, we calculate the force acting on the Casimir apparatus in the presence of the gravitational field. We first find the force acting on the single plate and then repeat the procedure for two parallel plates. For this, we first start with the general expression of the force density in curved space-time given by
\be
f^{\nu}=-\nabla_{\mu}T^{\mu \nu} =
 -\frac{\partial T^{\mu \nu}}{\partial x^{\mu}}-\Gamma^{\mu}_{\mu \sigma}T^{\sigma \nu}-\Gamma^{\nu}_{\mu \sigma}T^{\mu \sigma}.\label{L43}
\ee
Here $\nabla_{\mu}$ is the covariant derivative and $\partial_{\mu}=\frac{\partial}{\partial x^{\mu}}$ is the partial derivative. Generalizing the identity $\Gamma^{\mu}_{\mu \sigma}=\frac{1}{\sqrt{-g}}\partial_{\sigma}(\sqrt{-g})$ and properties of the Christoffel symbol \cite{shajesh2a,shajesh2b,moller} to the  $\kappa$-deformed Rindler space-time, we find force density expression for our study as
\be
\hat{f}_{\lambda}= -\frac{1}{\sqrt{-\hat{g}}}\partial_{\mu}(\sqrt{-\hat{g}} ~\hat{T}^{\mu}_{~~\lambda})+\frac{1}{2}\hat{T}^{\mu \sigma}(\partial_{\lambda}\hat{g}_{\mu \sigma}).\label{L44}
\ee
Using the expression of $\kappa$-deformed metric given in Eq.(\ref{N10}) in the above expression we find force density in Rindler coordinates as
\be
 \hat{f}_{\xi}=-(1+ap^{0})\Big(\frac{1}{\xi}\hat{T}_{\xi \xi}+\partial_{\xi}(\tilde{T}_{\xi \xi})\Big)-\frac{1}{\xi^{3}}\hat{T}_{0 0}.\label{L45}
\ee
In the above equation $\hat{T}_{\xi \xi}$ has dimension of $\text{Length}^{-4}$  and dimension of $\hat{T_{00}}$ is $\text{Length}^{-2}$.
Next from the expression of the energy-momentum tensor given in  Eq.(\ref{L13}) we find
\begin{multline}
\hat{T}_{00}=\frac{1}{2}\bigg\{(\partial_{\tau}\phi)^{2}+\xi^{2}\Big((\partial_{\xi}\phi)^{2}+(\partial_{x}\phi)^{2}+(\partial_{y}\phi)^{2}+V\phi^{2}\Big)\bigg\}\\+\frac{ap^{0}}{2}\xi^{2}\Big((\partial_{\xi}\phi)^{2}+(\partial_{x}\phi)^{2}+(\partial_{y}\phi)^{2}\Big)-\frac{a}{4}\Big(p^{1}\partial_{\tau}\phi\partial_{x}\phi+p^{2}\partial_{\tau}\phi\partial_{y}\phi+p^{3}\partial_{\tau}\phi\partial_{\xi}\phi\Big)\label{L46}
\end{multline}
and 
\begin{multline}
\hat{T}_{\xi \xi}=\frac{1}{2}\bigg\{\frac{1}{\xi^{2}}(\partial_{\tau}\phi)^{2}+(\partial_{\xi}\phi)^{2}-\Big((\partial_{x}\phi)^{2}+(\partial_{y}\phi)^{2}\Big)-V\phi^{2}\bigg\}\\-\frac{ap^{0}}{2}\Big(\frac{1}{\xi^{2}}(\partial_{\tau}\phi)^{2}-V\phi^{2}\Big)+\frac{a}{4\xi^{2}}\Big(p^{1}\partial_{\tau}\phi\partial_{x}\phi+p^{2}\partial_{\tau}\phi\partial_{y}\phi+p^{3}\partial_{\tau}\phi\partial_{\xi}\phi\Big).\label{L47}
\end{multline}
Using the above two equations in Eq.(\ref{L45}), and after re-expressing some of the terms using Eq.(\ref{L10}), we find
\bea
 \hat{f}_{\xi}&=&\frac{1}{2}\phi^{2}\partial_{\xi}V-\frac{1}{\xi^{2}}\partial_{\tau}(\partial_{\tau}\phi \partial_{\xi}\phi)+\nabla_{\perp}.(\nabla \phi \partial_{\xi}\phi) +ap^{0}\Big(\nabla_{\perp}.(\nabla \phi \partial_{\xi}\phi)\Big)\nonumber\\&&+ap^{1}\Big(\frac{1}{2\xi^{3}}\partial_{\tau}\phi \partial_{x}\phi-\frac{1}{4\xi^{2}}\partial_{\xi}(\partial_{\tau}\phi \partial_{x}\phi)-\frac{1}{2\xi^{2}}\partial_{\xi}\phi(\partial_{\tau}\partial_{x}\phi)\Big)\nonumber \\&&+ ap^{2}\Big(\frac{1}{2\xi^{3}}\partial_{\tau}\phi \partial_{y}\phi-\frac{1}{4\xi^{2}}\partial_{\xi}(\partial_{\tau}\phi \partial_{y}\phi)-\frac{1}{2\xi^{2}}\partial_{\xi}\phi(\partial_{\tau}\partial_{y}\phi)\Big) \nonumber \\&&+ ap^{3}\Big(\frac{1}{2\xi^{3}}\partial_{\tau}\phi \partial_{\xi}\phi-\frac{1}{4\xi^{2}}\partial_{\xi}(\partial_{\tau}\phi \partial_{\xi}\phi)-\frac{1}{2\xi^{2}}\partial_{\xi}\phi(\partial_{\tau}\partial_{\xi}\phi)\Big). \label{L48}
\eea
Next, we take the vacuum expectation value of the above  force density expression, and using it we find the total change in the momentum  of the Casimir apparatus in $\xi$ direction as
\bea
\Delta P_{\xi}&=&\int d\tau dx dy d\xi \sqrt{-\hat{g}}<\hat{f}_{\xi}> \nonumber\\
&=& \frac{1}{i}\Big(1-\frac{3}{2}ap^{0}\Big)\int d\tau dx dy d\xi \Bigg[\frac{1}{2}\xi(\partial_{\xi}V)\tilde{G}(x,x^{\prime})\Big\vert_{x=x^{\prime}}-\frac{1}{\xi}\partial_{\tau}\Big(\partial_{\tau}^{\prime}\partial_{\xi}\tilde{G}(x,x^{\prime})\Big\vert_{x=x^{\prime}}\Big)\nonumber \\&&+\xi \nabla_{\perp}.\Big(\nabla_{\perp}^{\prime}\partial_{\xi}\tilde{G}(x,x^{\prime})\Big\vert_{x=x^{\prime}}\Big) 
+ap^{0}\bigg\{\xi \nabla_{\perp}.\Big(\nabla_{\perp}^{\prime}\partial_{\xi}\tilde{G}(x,x^{\prime})\Big\vert_{x=x^{\prime}}\Big) \bigg\}+ap^{1}\xi\bigg\{\frac{1}{2\xi^{3}}\Big(\partial_{\tau}^{\prime}\partial_{x}\tilde{G}(x,x^{\prime})\Big\vert_{x=x^{\prime}}\Big)\nonumber \\&&-\frac{1}{4\xi^{2}}\partial_{\xi}\Big(\partial_{\tau}^{\prime}\partial_{x}\tilde{G}(x,x^{\prime})\Big\vert_{x=x^{\prime}}\Big)-\frac{1}{2\xi^{2}}\Big(\partial_{\xi}^{\prime} \partial_{\tau}\partial_{x}\tilde{G}(x,x^{\prime})\Big\vert_{x=x^{\prime}}\bigg\}+ ap^{2}\xi\bigg\{\frac{1}{2\xi^{3}}\Big(\partial_{\tau}^{\prime}\partial_{y}\tilde{G}(x,x^{\prime})\Big\vert_{x=x^{\prime}}\Big)\nonumber \\&&-\frac{1}{4\xi^{2}}\partial_{\xi}\Big(\partial_{\tau}^{\prime}\partial_{y}\tilde{G}(x,x^{\prime})\Big\vert_{x=x^{\prime}}\Big)-\frac{1}{2\xi^{2}}\Big(\partial_{\xi}^{\prime} \partial_{\tau}\partial_{y}\tilde{G}(x,x^{\prime})\Big\vert_{x=x^{\prime}}\bigg\}+ ap^{3}\xi\bigg\{\frac{1}{2\xi^{3}}\Big(\partial_{\tau}^{\prime}\partial_{\xi}\tilde{G}(x,x^{\prime})\Big\vert_{x=x^{\prime}}\Big)\nonumber \\&&-\frac{1}{4\xi^{2}}\partial_{\xi}\Big(\partial_{\tau}^{\prime}\partial_{\xi}\tilde{G}(x,x^{\prime})\Big\vert_{x=x^{\prime}}\Big)-\frac{1}{2\xi^{2}}\Big(\partial_{\xi}^{\prime} \partial_{\tau}\partial_{\xi}\tilde{G}(x,x^{\prime})\Big\vert_{x=x^{\prime}}\bigg\} \Bigg], \label{L49}
\eea
where we have used $\langle\phi(x^{\prime})\phi(x)\rangle=\frac{1}{i}\tilde{G}(x,x^{\prime},a)$. Next we use Fourier transformation of Green's function $\tilde{G}(x,x^{\prime},a)$ (see Eq.(\ref{L16})) and  Eq.(\ref{L18}) in above equation and after some simplification we find
total change in the momentum as
\bea
\Delta P_{\xi}&=& \frac{1}{i}\Big(1-\frac{3}{2}ap^{0}\Big)\int d\tau dx dy d\xi \int \frac{d\omega d^{2}k_{\perp}}{8\pi^{3}} \Bigg[\frac{1}{2}\partial_{\xi}\Big(\xi V \tilde{g}(\xi,\xi, a)\Big)-\frac{1}{2}V\partial_{\xi}\Big(\xi \tilde{g}(\xi,\xi, a)\Big) \nonumber \\&&
+ap^{1}\bigg\{\frac{1}{2\xi^{2}}(-\omega k_{x})\tilde{g}^{(0)}(\xi,\xi)-\frac{1}{4\xi}(-\omega k_{x})\partial_{\xi}\tilde{g}^{(0)}(\xi,\xi)-\frac{1}{2\xi}(\omega k_{x})\Big(\partial_{\xi}^{\prime}\tilde{g}^{(0)}(\xi,\xi^{\prime})\Big)\Big\vert_{\xi=\xi^{\prime}}\bigg\}\nonumber \\&&+ ap^{2}\bigg\{\frac{1}{2\xi^{2}}(-\omega k_{x})\tilde{g}^{(0)}(\xi,\xi)-\frac{1}{4\xi}(-\omega k_{x})\partial_{\xi}\tilde{g}^{(0)}(\xi,\xi)-\frac{1}{2\xi}(\omega k_{x})\Big(\partial_{\xi}^{\prime}\tilde{g}^{(0)}(\xi,\xi^{\prime})\Big)\Big\vert_{\xi=\xi^{\prime}}\bigg\}\nonumber \\&& + ap^{3}\bigg\{\frac{1}{2\xi^{2}}(i \omega)\Big(\partial_{\xi}\tilde{g}^{(0)}(\xi,\xi^{\prime})\Big)\Big\vert_{\xi=\xi^{\prime}}-\frac{1}{4\xi}(i \omega)\partial_{\xi}\Big(\partial_{\xi}\tilde{g}^{(0)}(\xi,\xi^{\prime})\Big\vert_{\xi=\xi^{\prime}}\Big)-\frac{1}{2\xi}(-i \omega)\Big(\partial_{\xi}^{\prime} \partial_{\xi}\tilde{g}^{(0)}(\xi,\xi^{\prime})\Big\vert_{\xi=\xi^{\prime}}\Big)\bigg\} \Bigg].\nonumber \\ \label{L50}
\eea
From the above equation, we find the general form of the force per unit area acting on the Casimir apparatus (single and two parallel plates) with the redefinition of angular frequency\footnote{Using $\omega=i\zeta$ we are canceling off the $i$ factor in the Eq.(\ref{L49}).}, $\omega=i\zeta$ to be, (at some arbitrary point $\xi_{0}$,)
\bea
F_{\xi}&=&\frac{\Delta P_{\xi}}{\xi_{0}\int d \tau dx dy}\nonumber \\ &=& \frac{1}{\xi_{0}}\Big(1-\frac{3}{2}ap^{0}\Big) \int \frac{ d\xi d\zeta d^{2}k_{\perp}}{8\pi^{3}} \Bigg[-\frac{1}{2}V\partial_{\xi}\Big(\xi \tilde{g}(\xi,\xi, a)\Big) \nonumber \\&&
+ap^{1}\bigg\{\frac{1}{2\xi^{2}}(-i\zeta k_{x})\tilde{g}^{(0)}(\xi,\xi)-\frac{1}{4\xi}(-i\zeta k_{x})\partial_{\xi}\tilde{g}^{(0)}(\xi,\xi)-\frac{1}{2\xi}(i\zeta k_{x})\Big(\partial_{\xi}^{\prime}\tilde{g}^{(0)}(\xi,\xi^{\prime})\Big)\Big\vert_{\xi=\xi^{\prime}}\bigg\}\nonumber \\&&+ ap^{2}\bigg\{\frac{1}{2\xi^{2}}(-i\zeta k_{x})\tilde{g}^{(0)}(\xi,\xi)-\frac{1}{4\xi}(- i\zeta k_{x})\partial_{\xi}\tilde{g}^{(0)}(\xi,\xi)-\frac{1}{2\xi}(i\zeta k_{x})\Big(\partial_{\xi}^{\prime}\tilde{g}^{(0)}(\xi,\xi^{\prime})\Big)\Big\vert_{\xi=\xi^{\prime}}\bigg\}\nonumber \\&& + ap^{3}\bigg\{\frac{1}{2\xi^{2}}(-\zeta)\Big(\partial_{\xi}\tilde{g}^{(0)}(\xi,\xi^{\prime})\Big)\Big\vert_{\xi=\xi^{\prime}}-\frac{1}{4\xi}(-\zeta)\partial_{\xi}\Big(\partial_{\xi}\tilde{g}^{(0)}(\xi,\xi^{\prime})\Big\vert_{\xi=\xi^{\prime}}\Big)-\frac{1}{2\xi}(\zeta)\Big(\partial_{\xi}^{\prime} \partial_{\xi}\tilde{g}^{(0)}(\xi,\xi^{\prime})\Big\vert_{\xi=\xi^{\prime}}\Big)\bigg\} \Bigg].\nonumber \\ \label{L51}
\eea
Next, we find the total force per unit area acting on the plates for two different cases, that is, a single plate and two parallel plates.
\subsection{Force acting on the single plate}
In this subsection, we find the total force acting on the single plate in the Minkowski-space limit (weak field approximation). We substitute $\tilde{g}(\xi,\xi, a)$ (see Eq.(\ref{L24})) in Eq.(\ref{L51}) and find the total force acting on the single plate to be
\be
F_{\xi}=F^{1}_{\xi}+F^{2}_{\xi},  \label{L52}
\ee
where 
\bea
F^{1}_{\xi}&=&\frac{(1-ap^{0})}{\xi_{0}}\int \frac{d\xi d\zeta d^{2}k_{\perp}}{8 \pi^{3}} \bigg\{-\frac{1}{2}\lambda_{1}\delta(\xi-\xi_{1})\partial_{\xi}\Big(\xi \tilde{g}^{(0)}(\xi,\xi)\Big)\bigg\}\nonumber \\ 
&=& \frac{(1-ap^{0})}{\xi_{0}}\int \frac{d\zeta d^{2}k_{\perp}}{8 \pi^{3}} \bigg\{-\frac{1}{2}\lambda_{1}\partial_{\xi_{1}}\Big(\xi_{1} \tilde{g}^{(0)}(\xi_{1},\xi_{1})\Big)\bigg\}\nonumber \\ 
&=& \frac{(1-ap^{0})}{\xi_{0}}\int \frac{d\zeta d^{2}k_{\perp}}{8 \pi^{3}} \bigg[-\frac{1}{2}\lambda_{1}\Big\{\partial_{\xi}\Big(\xi \tilde{g}^{(0)}(\xi,\xi)\Big)\Big\}_{\xi=\xi_{1}}\bigg] \label{L53}
\eea
and 
\bea
F^{2}_{\xi}&=&\frac{a}{\xi_{0}}\int \frac{d\xi d\zeta d^{2}k_{\perp}}{8 \pi^{3}} \Bigg[p^{1}\bigg\{\frac{1}{2\xi^{2}}(-i\zeta k_{x})\tilde{g}^{(0)}(\xi,\xi)-\frac{1}{4\xi}(-i\zeta k_{x})\partial_{\xi}\tilde{g}^{(0)}(\xi,\xi)-\frac{1}{2\xi}(i\zeta k_{x})\Big(\partial_{\xi}^{\prime}\tilde{g}^{(0)}(\xi,\xi^{\prime})\Big)\Big\vert_{\xi=\xi^{\prime}}\bigg\}\nonumber \\&&+ p^{2}\bigg\{\frac{1}{2\xi^{2}}(-i\zeta k_{x})\tilde{g}^{(0)}(\xi,\xi)-\frac{1}{4\xi}(- i\zeta k_{x})\partial_{\xi}\tilde{g}^{(0)}(\xi,\xi)-\frac{1}{2\xi}(i\zeta k_{x})\Big(\partial_{\xi}^{\prime}\tilde{g}^{(0)}(\xi,\xi^{\prime})\Big)\Big\vert_{\xi=\xi^{\prime}}\bigg\}\nonumber \\&& + p^{3}\bigg\{\frac{1}{2\xi^{2}}(-\zeta)\Big(\partial_{\xi}\tilde{g}^{(0)}(\xi,\xi^{\prime})\Big)\Big\vert_{\xi=\xi^{\prime}}-\frac{1}{4\xi}(-\zeta)\partial_{\xi}\Big(\partial_{\xi}\tilde{g}^{(0)}(\xi,\xi^{\prime})\Big\vert_{\xi=\xi^{\prime}}\Big)-\frac{1}{2\xi}(\zeta)\Big(\partial_{\xi}^{\prime} \partial_{\xi}\tilde{g}^{(0)}(\xi,\xi^{\prime})\Big\vert_{\xi=\xi^{\prime}}\Big)\bigg\} \Bigg].\nonumber \\ \label{L54}
\eea
In Eq.(\ref{L53}), the first derivative of $\tilde{g}^{(0)}(\xi,\xi)$ is discontinious at $\xi=\xi_{1}$, which is a jump discontiunity. Interpreting the value of a function at the point where it has jump discontinuity to be the average of jump leads us to evaluate 
\be
 \Big\{\xi\partial_{\xi}\tilde{g}^{(0)}(\xi,\xi)\Big\}_{\xi=\xi_{1}}=\frac{1}{2}\bigg[\Big\{\xi \partial_{\xi}\tilde{g}^{(0)}(\xi,\xi)\Big\}_{\xi=\xi^{+}_{1}}+\Big\{\xi \partial_{\xi}\tilde{g}^{(0)}(\xi,\xi)\Big\}_{\xi=\xi^{-}_{1}}\bigg]. \label{L55}
\ee
Using Eq.(\ref{L21}) and Eq.(\ref{L22}) in above equation we find
\be
\Big\{\xi\partial_{\xi}\tilde{g}^{(0)}(\xi,\xi)\Big\}_{\xi=\xi_{1}}=\frac{\xi_{1}}{1+\lambda_{1}\xi_{1}I_{\zeta}(k_{\perp} \xi_{1})K_{\zeta}(k_{\perp} \xi_{1})}\partial_{\xi_{1}}\Big(I_{\zeta}(k_{\perp} \xi_{1})K_{\zeta}(k_{\perp} \xi_{1})\Big). \label{L56}
\ee
Using this, we evaluate the right-hand side of Eq.(\ref{L53}), the term in the square bracket, to be 
\be
 \lambda_{1}\bigg\{\partial_{\xi}\Big(\xi\tilde{g}^{(0)}(\xi,\xi)\Big)\bigg\}_{\xi=\xi_{1}} = \partial_{\xi_{1}}\text{ln}\Big(1+\lambda_{1}\xi_{1}I_{\zeta}(k_{\perp} \xi_{1})K_{\zeta}(k_{\perp} \xi_{1})\Big).\label{L57}
\ee
Substituting this in Eq.(\ref{L53}) we find
\be
 F^{1}_{\xi}=-\frac{(1-ap^{0})}{2\xi_{0}}\frac{\partial}{\partial \xi_{1}}\int \frac{d\zeta d^{2}k_{\perp}}{8 \pi^{3}}\text{ln}\Big(1+\lambda_{1}\xi_{1}I_{\zeta}(k_{\perp} \xi_{1})K_{\zeta}(k_{\perp} \xi_{1})\Big).  \label{L58}
\ee
Next we take the Minkowski-space limit (see Eq.(\ref{L25}) to Eq.(\ref{L28})) of the above equation\footnote{We use Eq.(\ref{L26}), Eq.(\ref{L27}) and Eq.(\ref{L28}) for taking the asympototic expansion about the arbitary point $\xi_{0}$ and find
\begin{equation*}
 \xi_{1}I_{\zeta}(k_{\perp}\xi_{1})K_{\zeta}(k_{\perp}\xi_{1}) \sim \frac{\xi_{1}}{2 \zeta}\frac{1}{\sqrt{1+\Big(\frac{k_{\perp}\xi_{1}}{\zeta}\Big)^{2}}} \approx \frac{1}{2 \bar{k}}\bigg(1+\frac{\xi_{1}-\xi_{0}}{\xi_{0}}\bigg)\bigg(1-\frac{k_{\perp}^{2}}{\bar{k}^{2}}\frac{(\xi_{1}-\xi_{0})}{\xi_{0}}\bigg).
 \end{equation*}
 Note that in finding the above expression, we have considered $\xi_{1} \approx \xi_{0}$, that is, $\xi_{1}+\xi_{2} \approx 2\xi_{0}$.}
 and we find
 \bea
 F^{1}_{\xi}&=&-(1-ap^{0})\frac{\lambda_{1}}{16 \pi^{3} \xi_{0}}\int  \frac{d\zeta d^{2}k_{\perp}}{2\bar{k}+\lambda_{1}\bigg(1+\frac{\xi_{1}-\xi_{0}}{\xi_{0}}\bigg)\bigg(1-\frac{k_{\perp}^{2}}{\bar{k}^{2}}\frac{(\xi_{1}-\xi_{0})}{\xi_{0}}\bigg)} \partial_{\xi_{1}}\Bigg\{ \bigg(1+\frac{\xi_{1}-\xi_{0}}{\xi_{0}}\bigg)\bigg(1-\frac{k_{\perp}^{2}}{\bar{k}^{2}}\frac{(\xi_{1}-\xi_{0})}{\xi_{0}}\bigg)\Bigg\}\nonumber \\
 &=&-(1-ap^{0})\frac{\lambda_{1}}{16 \pi^{3}\xi_{0}}\int  \frac{d\hat{\zeta} d^{2}k_{\perp}}{2\bar{k}+\lambda_{1}\bigg(1+\frac{\xi_{1}-\xi_{0}}{\xi_{0}}\bigg)\bigg(1-\frac{k_{\perp}^{2}}{\bar{k}^{2}}\frac{(\xi_{1}-\xi_{0})}{\xi_{0}}\bigg)} \frac{\hat{\zeta}^{2}}{\bar{k}^{2}}\partial_{\xi_{1}}(\xi_{1}-\xi_{0}).
  \label{L59}
\eea
Here $\hat{\zeta}=\frac{\zeta}{\xi_{0}}$.
Using spherical polar coordinates for phase space we re-express $\int^{+ \infty}_{-\infty} \frac{d \hat{\zeta} d^{2}k_{\perp}}{16 \pi^{3}}\frac{\hat{\zeta}^{2}}{\bar{k}^{2}}$ as $\frac{1}{12\pi^{2}}\int^{\infty}_{0} \bar{k}^{2} d\bar{k}$ and we find
\bea
 F^{1}_{\xi}&=& -(1-ap^{0})\frac{\lambda_{1}}{12 \pi^{2}\xi_{0}}\int^{\infty}_{0} \frac{\bar{k}^{2} d\bar{k}}{\lambda_{1}+2\bar{k}} \nonumber \\
 &=&-(1-ap^{0})\frac{1}{\xi_{0}}\frac{\lambda_{1}}{96 \pi^{2}}\int^{\infty}_{0}\frac{Y^{2} dY}{\lambda_{1}+Y},\label{L60}
\eea
where $Y=2\bar{k} $. We have taken $\xi_{1} \approx \xi_{0}$ for calculational simplification in the above equations. We observe that the above equation is divergent. Next we take Minkowski-space limit of Eq.(\ref{L54}) and using eq.(\ref{L29}) we find
\bea
F^{2}_{\xi}&=&\frac{a}{\xi_{0}}\int \frac{d\xi d\zeta d^{2}k_{\perp}}{8 \pi^{3}} \Bigg[p^{1}\bigg\{\frac{1}{2\xi_{0}^{3}}(-i\zeta k_{x})g^{(0)}(\xi,\xi)-\frac{1}{4\xi^{2}_{0}}(-i\zeta k_{x})\partial_{\xi}g^{(0)}(\xi,\xi)-\frac{1}{2\xi^{2}_{0}}(i\zeta k_{x})\Big(\partial_{\xi}^{\prime}g^{(0)}(\xi,\xi^{\prime})\Big)\Big\vert_{\xi=\xi^{\prime}}\bigg\}\nonumber \\&&+ p^{2}\bigg\{\frac{1}{2\xi_{0}^{3}}(-i\zeta k_{x})g^{(0)}(\xi,\xi)-\frac{1}{4\xi^{2}_{0}}(- i\zeta k_{x})\partial_{\xi}g^{(0)}(\xi,\xi)-\frac{1}{2\xi^{2}_{0}}(i\zeta k_{x})\Big(\partial_{\xi}^{\prime}g^{(0)}(\xi,\xi^{\prime})\Big)\Big\vert_{\xi=\xi^{\prime}}\bigg\}\nonumber \\&& + p^{3}\bigg\{\frac{1}{2\xi_{0}^{3}}(-\zeta)\Big(\partial_{\xi}g^{(0)}(\xi,\xi^{\prime})\Big)\Big\vert_{\xi=\xi^{\prime}}-\frac{1}{4\xi^{2}_{0}}(-\zeta)\partial_{\xi}\Big(\partial_{\xi}g^{(0)}(\xi,\xi^{\prime})\Big\vert_{\xi=\xi^{\prime}}\Big)-\frac{1}{2\xi^{2}_{0}}(\zeta)\Big(\partial_{\xi}^{\prime} \partial_{\xi}g^{(0)}(\xi,\xi^{\prime})\Big\vert_{\xi=\xi^{\prime}}\Big)\bigg\} \Bigg]\nonumber \\
&\approx & \frac{a}{\xi_{0}}\bigg\{\mathcal{O}\Big(\frac{1}{\xi_{0}}\Big)^{2}\bigg\}. \label{L61} 
\eea
Thus, using Eq.(\ref{L60}) and Eq.(\ref{L61}) in Eq.(\ref{L52}) and by ignoring the very small terms (i.e., $\mathcal{O}\Big(\frac{1}{\xi_{0}}\Big)^{2}$) we find the total force acting on the single plate as
\be
F_{\xi}= -(1-ap^{0})\frac{1}{96 \pi^{2}\xi_{0}}\int^{\infty}_{0}\frac{Y^{2} dY}{1+\frac{Y}{\lambda_{1}}}. \label{L62}
\ee
We see that the non-commutativity of space-time results in the modification of the total force by an overall multiplicative factor. Note here that we have taken correction terms valid only up to the first order in deformation parameter $a$. In the limit, $a \rightarrow 0$ above, the expression of force acting on the single plate reduces to the commutative result \cite{shajesh2a,shajesh2b}.

\subsection{Force acting on two parallel plates}
In this subsection, we find the total force acting on the two parallel plates situated at $\xi=\xi_{1}$ and $\xi=\xi_{2}$, respectively. Using Eq.(\ref{L34})  in Eq.(\ref{L51}) we find the total force, valid up to the first order in deformation parameter $a$ as
\be
F_{\xi}=F^{1}_{\xi}+F^{2}_{\xi}, \label{L63a}
\ee
where 
\bea
F^{1}_{\xi}&=& -\frac{(1-ap^{0})}{2\xi_{0}}\int \frac{d\zeta d^{2}k_{\perp}}{8 \pi^{3}} \bigg\{\lambda_{1}\partial_{\xi_{1}}\Big(\xi_{1} \tilde{g}^{(0)}(\xi_{1},\xi_{1})\Big)+\lambda_{2}\partial_{\xi_{2}}\Big(\xi_{2} \tilde{g}^{(0)}(\xi_{2},\xi_{2})\Big)\bigg\}\nonumber \\ 
&=& -\frac{(1-ap^{0})}{2\xi_{0}}\int \frac{d\zeta d^{2}k_{\perp}}{8 \pi^{3}} \bigg[\lambda_{1}\Big\{\partial_{\xi}\Big(\xi \tilde{g}^{(0)}(\xi,\xi)\Big)\Big\}_{\xi=\xi_{1}}+\lambda_{2}\Big\{\partial_{\xi}\Big(\xi \tilde{g}^{(0)}(\xi,\xi)\Big)\Big\}_{\xi=\xi_{2}}\bigg] \label{L63}
\eea
and 
\bea
F^{2}_{\xi}&=&\frac{a}{\xi_{0}}\int \frac{d\xi d\zeta d^{2}k_{\perp}}{8 \pi^{3}} \Bigg[p^{1}\bigg\{\frac{1}{2\xi^{2}}(-i\zeta k_{x})\tilde{g}^{(0)}(\xi,\xi)-\frac{1}{4\xi}(-i\zeta k_{x})\partial_{\xi}\tilde{g}^{(0)}(\xi,\xi)-\frac{1}{2\xi}(i\zeta k_{x})\Big(\partial_{\xi}^{\prime}\tilde{g}^{(0)}(\xi,\xi^{\prime})\Big)\Big\vert_{\xi=\xi^{\prime}}\bigg\}\nonumber \\&&+ p^{2}\bigg\{\frac{1}{2\xi^{2}}(-i\zeta k_{x})\tilde{g}^{(0)}(\xi,\xi)-\frac{1}{4\xi}(- i\zeta k_{x})\partial_{\xi}\tilde{g}^{(0)}(\xi,\xi)-\frac{1}{2\xi}(i\zeta k_{x})\Big(\partial_{\xi}^{\prime}\tilde{g}^{(0)}(\xi,\xi^{\prime})\Big)\Big\vert_{\xi=\xi^{\prime}}\bigg\}\nonumber \\&& + p^{3}\bigg\{\frac{1}{2\xi^{2}}(-\zeta)\Big(\partial_{\xi}\tilde{g}^{(0)}(\xi,\xi^{\prime})\Big)\Big\vert_{\xi=\xi^{\prime}}-\frac{1}{4\xi}(-\zeta)\partial_{\xi}\Big(\partial_{\xi}\tilde{g}^{(0)}(\xi,\xi^{\prime})\Big\vert_{\xi=\xi^{\prime}}\Big)-\frac{1}{2\xi}(\zeta)\Big(\partial_{\xi}^{\prime} \partial_{\xi}\tilde{g}^{(0)}(\xi,\xi^{\prime})\Big\vert_{\xi=\xi^{\prime}}\Big)\bigg\} \Bigg],\nonumber \\ \label{L64}
\eea
where expressions for $\tilde{g}^{(0)}(\xi,\xi^{\prime})$ are given in Eq.(\ref{L35}), Eq.(\ref{L36}) and Eq.(\ref{L37}) for three different regions. We observe here that  $\tilde{g}^{(0)}(\xi,\xi)$ is continuous at $\xi = \xi_{1}$ and $\xi =\xi_{2}$. Thus, using Eq.(\ref{L35}), Eq.(\ref{L36}) and Eq.(\ref{L37}) we find
\be
\tilde{g}^{(0)}(\xi_{1},\xi_{1})=\frac{1}{\tilde{\Delta}}\bigg\{I_{\zeta}(k_{\perp}\xi_{1})K_{\zeta}(k_{\perp}\xi_{1})+\lambda_{2}\xi_{2}\Big(I_{\zeta}(k_{\perp}\xi_{1})K_{\zeta}(k_{\perp}\xi_{1})I_{\zeta}(k_{\perp}\xi_{2})K_{\zeta}(k_{\perp}\xi_{2})-I_{\zeta}^{2}(k_{\perp}\xi_{1})K_{\zeta}^{2}(k_{\perp}\xi_{2})\Big)\bigg\} \label{L65}
\ee
 and 
\be
\tilde{g}^{(0)}(\xi_{2},\xi_{2})=\frac{1}{\tilde{\Delta}}\bigg\{I_{\zeta}(k_{\perp}\xi_{2})K_{\zeta}(k_{\perp}\xi_{2})+\lambda_{1}\xi_{1}\Big(I_{\zeta}(k_{\perp}\xi_{1})K_{\zeta}(k_{\perp}\xi_{1})I_{\zeta}(k_{\perp}\xi_{2})K_{\zeta}(k_{\perp}\xi_{2})-I_{\zeta}^{2}(k_{\perp}\xi_{1})K_{\zeta}^{2}(k_{\perp}\xi_{2})\Big)\bigg\}. \label{L66}
\ee
Again, first derivative of $\tilde{g}^{(0)}(\xi,\xi)$ is discontinuous at $\xi=\xi_{1}$ and $\xi=\xi_{2}$, that is, it has jump discontinuity at $\xi=\xi_{1}$
 and $\xi=\xi_{2}$. Thus, by following the same procedure discussed for single plate in the previous subsection and using Eq.(\ref{L35}) and Eq.(\ref{L36}) we find
 \begin{multline}
  \Big\{\xi \partial_{\xi}\Big( \tilde{g}^{(0)}(\xi,\xi)\Big)\Big\}_{\xi=\xi_{1}}= \frac{1}{\tilde{\Delta}}\Bigg[\xi_{1}\partial_{\xi_{1}}\bigg\{I_{\zeta}(k_{\perp}\xi_{1})K_{\zeta}(k_{\perp}\xi_{1})\\+\lambda_{2}\xi_{2}\Big(I_{\zeta}(k_{\perp}\xi_{1})K_{\zeta}(k_{\perp}\xi_{1})I_{\zeta}(k_{\perp}\xi_{2})K_{\zeta}(k_{\perp}\xi_{2})-I_{\zeta}^{2}(k_{\perp}\xi_{1})K_{\zeta}^{2}(k_{\perp}\xi_{2})\Big)\bigg\}\Bigg]. \label{L67}
 \end{multline}
Similarly using Eq.(\ref{L36}) and Eq.(\ref{L37}) we find
\begin{multline}
  \Big\{\xi \partial_{\xi}\Big( \tilde{g}^{(0)}(\xi,\xi)\Big)\Big\}_{\xi=\xi_{2}}= \frac{1}{\tilde{\Delta}}\Bigg[\xi_{2}\partial_{\xi_{2}}\bigg\{I_{\zeta}(k_{\perp}\xi_{2})K_{\zeta}(k_{\perp}\xi_{2})\\+\lambda_{1}\xi_{1}\Big(I_{\zeta}(k_{\perp}\xi_{1})K_{\zeta}(k_{\perp}\xi_{1})I_{\zeta}(k_{\perp}\xi_{2})K_{\zeta}(k_{\perp}\xi_{2})-I_{\zeta}^{2}(k_{\perp}\xi_{1})K_{\zeta}^{2}(k_{\perp}\xi_{2})\Big)\bigg\}\Bigg]. \label{L68}
 \end{multline}
Next using Eq.(\ref{L65}) and Eq.(\ref{L67}) we find
\bea
\lambda_{1}\Big\{ \partial_{\xi}\Big( \xi\tilde{g}^{(0)}(\xi,\xi)\Big)\Big\}_{\xi=\xi_{1}} &=&\frac{1}{\tilde{\Delta}}\Bigg[\partial_{\xi_{1}}\bigg\{1+\lambda_{1}\xi_{1}I_{\zeta}(k_{\perp}\xi_{1})K_{\zeta}(k_{\perp}\xi_{1})+\lambda_{2}\xi_{2}I_{\zeta}(k_{\perp}\xi_{2})K_{\zeta}(k_{\perp}\xi_{2})\nonumber \\&&+\lambda_{1}\xi_{1}\lambda_{2}\xi_{2}\Big(I_{\zeta}(k_{\perp}\xi_{1})K_{\zeta}(k_{\perp}\xi_{1})I_{\zeta}(k_{\perp}\xi_{2})K_{\zeta}(k_{\perp}\xi_{2})-I_{\zeta}^{2}(k_{\perp}\xi_{1})K_{\zeta}^{2}(k_{\perp}\xi_{2})\Big)\bigg\}\Bigg]\nonumber \\
&=& \frac{1}{\tilde{\Delta}}\partial_{\xi_{1}}\tilde{\Delta}. \label{L69}
\eea
Similarly using Eq.(\ref{L66}) and Eq.(\ref{L68}) we get
\be
 \lambda_{2}\Big\{ \partial_{\xi}\Big( \xi\tilde{g}^{(0)}(\xi,\xi)\Big)\Big\}_{\xi=\xi_{2}}=\frac{1}{\tilde{\Delta}}\partial_{\xi_{2}}\tilde{\Delta}. \label{L70}
\ee
Using Eq.(\ref{L69}) and Eq.(\ref{L70}) in Eq.(\ref{L63}) we obtain
\be
F^{1}_{\xi}= -\frac{(1-ap^{0})}{2\xi_{0}}\int \frac{d\zeta d^{2}k_{\perp}}{8 \pi^{3}} \Big(\partial_{\xi_{1}}(\text{ln} \tilde{\Delta})+\partial_{\xi_{2}}(\text{ln} \tilde{\Delta})\Big).\label{L71}
\ee
Next, we take the Minkowski-space limit  (see for details Eq.(\ref{L25}) to Eq.(\ref{L28}) and corresponding discussions) and find
\be
\tilde{\Delta}(\xi_{1},\xi_{2})= \Delta^{(0)}(\xi_{2}-\xi_{1})+\frac{1}{\xi_{0}}\Delta^{(1)}(\xi_{1},\xi_{2}) + \mathcal{O}\Big(\frac{1}{\xi_{0}}\Big)^{2},\label{L72}
\ee
where 
\be
\Delta^{(0)}=1+\frac{\lambda_{1}}{2\bar{k}}+\frac{\lambda_{2}}{2\bar{k}}+\frac{\lambda_{1}}{2\bar{k}}\frac{\lambda_{2}}{2\bar{k}}\Big(1-e^{-2\bar{k}L}\Big), \label{L73}
\ee
and 
\be
\Delta^{(1)}=\frac{\hat{\zeta}^{2}}{\bar{k}^{2}}\Big(\frac{\xi_{1}+\xi_{2}}{2}-\xi_{0}\Big)\Bigg[ \frac{\lambda_{1}}{2\bar{k}}+\frac{\lambda_{2}}{2\bar{k}}+2 \frac{\lambda_{1}}{2\bar{k}} \frac{\lambda_{2}}{2\bar{k}}\bigg\{1-e^{-2\bar{k}L}\Big(1+\bar{k}L\Big)\bigg\}\Bigg]. \label{L74}
\ee
Here in the above equations $L=(\xi_{2}-\xi_{1})$ is the distance between the plates. Using Eq.(\ref{L72}), Eq.(\ref{L73}) and Eq.(\ref{L74}) in Eq.(\ref{L71}) we find
\be
F^{1}_{\xi}= -\frac{(1-ap^{0})}{2\xi_{0}}\int \frac{d\hat{\zeta} d^{2}k_{\perp}}{8 \pi^{3}} \frac{\hat{\zeta}^{2}}{\bar{k}^{2}}\frac{\Bigg[ \frac{\lambda_{1}}{2\bar{k}}+\frac{\lambda_{2}}{2\bar{k}}+2 \frac{\lambda_{1}}{2\bar{k}} \frac{\lambda_{2}}{2\bar{k}}\bigg\{1-e^{-2\bar{k}L}\Big(1+\bar{k}L\Big)\bigg\}\Bigg]}{\bigg[1+\frac{\lambda_{1}}{2\bar{k}}+\frac{\lambda_{2}}{2\bar{k}}+\frac{\lambda_{1}}{2\bar{k}}\frac{\lambda_{2}}{2\bar{k}}\Big(1-e^{-2\bar{k}L}\Big)\bigg]}.\label{L75}
\ee
Next using spherical polar coordinate for phase space (see discussion after Eq.(\ref{L59})) we find
%\footnote{ Here we have used $\int^{+ \infty}_{-\infty} \frac{d \hat{\zeta} d^{2}k_{\perp}}{16 \pi^{3}}\frac{\hat{\zeta}^{2}}{\bar{k}^{2}} = \frac{1}{12\pi^{2}}\int^{\infty}_{0} \bar{k}^{2} d\bar{k}$.} 
\be
F^{1}_{\xi}=-(1-ap^{0})\frac{1}{96\pi^{2} L^{3}\xi_{0}}\int^{\infty}_{0}dY Y^{2}\frac{\bigg[\frac{Y}{\lambda_{1}L}+\frac{Y}{\lambda_{2}L}+2\Big\{1-e^{-Y}\Big(1+\frac{Y}{2}\Big)\Big\}\bigg]}{\bigg[\frac{Y^{2}}{\lambda_{1}\lambda_{2}L^{2}}+\frac{Y}{\lambda_{1}L}+\frac{Y}{\lambda_{2}L}+\Big(1-e^{-Y}\Big)\bigg]};~~\text{where}~~Y=2\bar{k}L. \label{L76}
\ee
Which we re-express as
\begin{multline}
 F^{1}_{\xi}=-(1-ap^{0})\frac{1}{96\pi^{2} L^{3}\xi_{0}}\Bigg[\int^{\infty}_{0}dY Y^{2}\frac{1}{\Big(1+\frac{Y}{\lambda_{1}L}\Big)}+\int^{\infty}_{0}dY Y^{2}\frac{1}{\Big(1+\frac{Y}{\lambda_{1}L}\Big)} \\-\int^{\infty}_{0}dY Y^{3}\frac{1+\frac{1}{Y+\lambda_{1}L} +\frac{1}{Y+\lambda_{2}L}}{\Big(1\frac{Y}{\lambda_{1}L}\Big)\Big(1+\frac{Y}{\lambda_{2}L}\Big)e^{Y}-1}\Bigg]. \label{L77}
\end{multline}
As in the case of the single plate discussed in the previous subsection, here also we take the Minkowski-space limit (weak field approximation) and find $F^{2}_{\xi} \approx  a~\mathcal{O}\Big(\frac{1}{\xi_{0}}\Big)^{2}$. Since contribution from $\mathcal{O}\Big(\frac{1}{\xi_{0}}\Big)^{2}$ dependent terms are negligible, from Eq.(\ref{L63a}) we find the total force acting on the two parallel plates as
\bea
 F_{\xi}&=&-(1-ap^{0})\frac{1}{96\pi^{2} L^{3}\xi_{0}}\Bigg[\int^{\infty}_{0}dY Y^{2}\frac{1}{\Big(1+\frac{Y}{\lambda_{1}L}\Big)}+\int^{\infty}_{0}dY Y^{2}\frac{1}{\Big(1+\frac{Y}{\lambda_{1}L}\Big)} \nonumber \\&&-\int^{\infty}_{0}dY Y^{3}\frac{1+\frac{1}{Y+\lambda_{1}L} +\frac{1}{Y+\lambda_{2}L}}{\Big(1\frac{Y}{\lambda_{1}L}\Big)\Big(1+\frac{Y}{\lambda_{2}L}\Big)e^{Y}-1}\Bigg]  \nonumber \\
 &=&-\frac{(\tilde{E}_{d1}+\tilde{E}_{d2}+\tilde{E}_{c})}{\xi_{0}}, \label{L78}
\eea
where $\tilde{E}_{d1}$ and $\tilde{E}_{d2}$ are the divergent energies corresponding to the two parallel plates and $\tilde{E}_{c}$ is the Casimir energy of the system of two semitransparent parallel plates in the $\kappa$-deformed space-time (for details see Appendix-A)
\be
\tilde{E}_{c} = -(1-ap^{0})\frac{1}{96\pi^{2} L^{3}}\int^{\infty}_{0}dY Y^{3}\frac{1+\frac{1}{Y+\lambda_{1}L} +\frac{1}{Y+\lambda_{2}L}}{\Big(1+\frac{Y}{\lambda_{1}L}\Big)\Big(1+\frac{Y}{\lambda_{2}L}\Big)e^{Y}-1}.\label{A23a}
\ee
Here, too, we find that the non-commutative corrections to the force appear through an overall multiplicative factor. We observe that the total force acting on the two parallel plates in $\kappa$-deformed Rindler space-time is equal to the total energy associated with the two parallel plates in $\kappa$-deformed Minkowski space-time multiplied by the acceleration of the Casimir apparatus (that is, $\frac{1}{\xi_{0}}$) (See Appendix-A). Next, we use mass renormalization (per unit area) of each plate (divergent terms in Eq.(\ref{L78})) and find
\be
\tilde{E}_{\text{total}}=\tilde{E}_{d1}+m_{1}+\tilde{E}_{d2}+m_{2}+\tilde{E}_{c} =\tilde{M}_{1}+\tilde{M}_{2}+\tilde{E}_{c}, \label{L79}
\ee
where $\tilde{M}_{1}$ and $\tilde{M}_{2}$ are the renormalized masses for the plates and $m_{1}$ and $m_{2}$ are the bare mass of the plates. Thus, using Eq.(\ref{L79}) in Eq.(\ref{L78}) we find the gravitational force acting on the Casimir apparatus to be
\be
F_{\xi}=- (\tilde{M}_{1}+\tilde{M}_{2}+\tilde{E}_{c})\frac{1}{\xi_{0}}=-(\tilde{M}_{1}+\tilde{M}_{2}+\tilde{E}_{c})g.  \label{L80}
\ee
Note here that the negative sign in the above equation indicates the downward acceleration of gravity and we have set $\frac{1}{\xi_{0}}=g$. This shows that the  Casimir energy (two semitransparent parallel plates) falls in the presence of a gravitational field and thus satisfies the mass-energy equivalence principle. We observe that in Eq.(\ref{L78}) Casimir energy is attractive, which reduces the gravitational force acting on the Casimir apparatus. It is important to note that correction terms, which are coming due to the non-commutativity of space-time, reduce the gravitational force acting on the two parallel plates. Nevertheless, we observe that kappa-deformed Casimir energy gravitates in accordance with the mass-energy equivalence principle. Note that the Casimir energy $\tilde{E}_{c}$ in Eq.(\ref{A23a}) depends on the non-commutative parameter $a$. The presence of the parameter $a$, having length dimension in $\tilde{E}_{c}$ shows that even in the weak gravity regime that we are analysing the effect of non-commutativity of space-time is relavent. The existance of a fundamental length scale (which is a hallmark of quantum gravity) does not affect the validity of equivalance principle. In all the above calculations we included correction terms valid up to the first order in the deformation parameter $a$. Note that the non-commutative dependent modification is always appearing through the factor $ap^{0}$. Here $p^{0}$ is the energy of the scalar quanta that is producing the Casimir energy. Thus, we see that scalar quanta with different energies are accelerated by different amounts. The above expression of gravitational force acting on the Casimir apparatus will reduce to the commutative result \cite{shajesh2a,shajesh2b} in the limit $a \rightarrow 0$.

\section{Conclusions}
In this study, we investigated the effects of the non-commutativity of space-time on the free fall of the Casimir energy in the presence of a weak gravitational field. For this we considered the Casimir apparatus moving with constant acceleration. This setup
is modeled using $\kappa$-deformed Rindler coordinates. We note from Eq.(\ref{khyp}) that the particle moving in the $\kappa$-deformed Rindler space-time takes less time to travel the same distance compared to the particle moving in the commutative counterpart.

We calculate the Casimir force experienced by the plates (single and double plate configurations) in the presence of $\kappa$-deformed Klein-Gordon field. All our calculations are valid up to the first order in the deformation parameter. For this, first, we construct $\kappa$-deformed Klein-Gordon Lagrangian in the presence of the Casimir plates. Using this, one derives the $\kappa$-deformed Energy-Momentum tensor. The vacuum expectation value of the Energy-Momentum tensor is related to Green's function, and for this reason, we first derive the $\kappa$-deformed Green's function, valid up to the first order in the deformation parameter, corresponding to the $\kappa$-deformed Klein-Gordon field. From the $\kappa$-deformed Energy-Momentum tensor, one finds the force experienced by the Casimir apparatus-i.e., the $\kappa$-deformed Casimir force. Comparing the $\kappa$-deformed Casimir energy for a single plate and $\kappa$-deformed force on a double plate, we show that the $\kappa$-deformed Casimir force on the double plate configuration is exactly the same as the sum of the deformed Casimir energy and renormalized masses of the plates multiplied by the acceleration due to weak gravity. This proves the validity of the mass-energy equivalence principle in $\kappa$-deformed space-time.

We note from Eq.(\ref{A23}) that the magnitude of the Casimir energy in $\kappa$-space-time is lower than that in the commutative space-time. Further, the non-commutative correction depends on $ap^{0}$ where $a$ is the fundamental length scale introduced by the non-commutativity of the space-time and $p^{0}$ is the energy of the probe that reveals the non-commutative structure of the space-time. This clearly shows that even in the non-commutative space-time, the equivalence principle is valid.
%(Since for $a \neq 0$, $p^{0}$ can take any value, this interpretation is more appropriate rather than taking $p^{0}$ to be the energy scale at which space-time becomes non-commutative).

Our study shows that in the $\kappa$-deformed space-time, Casimir energy is accelerated downward in the weak gravitational field. Thus, we see that the Casimir energy in the $\kappa$-deformed space-time respects the mass-energy equivalence principle, as in the commutative space-time \cite{shajesh2a,shajesh2b}. In the limit of commutative space-time, i.e., $a \rightarrow 0$, we recover the results obtained in \cite{shajesh2a,shajesh2b}. The correction to the Casimir energy due to non-commutativity comes with an opposite sign compared to the Casimir energy in the commutative space-time. Thus, the net downward force experienced by the Casimir apparatus in the $\kappa$-deformed space-time is lower than that in the commutative space-time. The amount of this reduction in the force depends on $ap^{0}$. Thus, for $a \neq 0$, the magnitude of the reduction in the downward gravitational force experienced by the Casimir apparatus depends on the energy of the probe. Since the acceleration of the $\kappa$-deformed Rindler observer is $\frac{1}{\xi}$, which is the same as that in the commutative space-time, `$g$' appearing in Eq.(\ref{L80}) is same as that of the commutative space-time. Thus, it is natural to conclude that the equivalence principle holds in the $\kappa$-deformed space-time.

\renewcommand{\thesection}{Appendix-A}
\section{Total energy of the Casimir apperatus in $\kappa$-deformed minkowski space-time}
\renewcommand{\thesection}{A}
Here we find the total energy associated with the Casimir apparatus (two parallel plates)  in the $\kappa$- deformed Minkowski space-time. For this first we construct the Lagrangian by the procedure discussed in Sec. (4) (for details see Eq.(\ref{L1}) to Eq.(\ref{L8})). Thus we find the expression of the Lagrangian for scalar field interacting with two parallel plates situated at $z=z_{1}$ and $z=z_{2}$ as
\bea
 L&=&-\frac{1}{2}\eta^{(0){\mu\nu}}\partial_{\mu}\phi \partial_{\nu}\phi-\frac{1}{2}\eta^{(1){\mu\nu}}(a)\partial_{\mu}\phi \partial_{\nu}\phi \nonumber -\frac{1}{2}V\phi^{2}\\
 &=& \frac{1}{2}(\partial_{0}\phi)^{2}-\frac{1}{2}\Big((\partial_{x}\phi)^{2}+(\partial_{y}\phi)^{2}+(\partial_{z}\phi)^{2}\Big) -\frac{1}{2}V\phi^{2}\nonumber \\&&-\frac{ap^{0}}{2}\Big((\partial_{x}\phi)^{2}+(\partial_{y}\phi)^{2}+(\partial_{z}\phi)^{2}\Big)+\frac{ap^{1}}{4}\partial_{0}\phi\partial_{x}\phi + \frac{ap^{2}}{4}\partial_{0}\phi\partial_{y}\phi + \frac{ap^{3}}{4}\partial_{0}\phi\partial_{z}\phi \label{A1}
\eea
where 
\be
V(x)= \lambda_{1}\delta(z-z_{1})+\lambda_{2}\delta(z-z_{2}).\label{A2}
\ee
In the above equation $\eta^{(0){\mu\nu}}$ and $\eta^{(1){\mu\nu}}(a)$  are the Minkowski metric and it's $\kappa$-deformed correction. We find $\kappa$-deformed Minkowski metric up to first order in deformation parameter $a$ using Eq.(\ref{N9}) as
\bea
\hat{\eta}_{\mu\nu}(a) &=& \eta^{(0)}_{\mu\nu}+\eta^{(1)}_{\mu\nu}(a) \nonumber \\ 
&=&\left[\begin{array}{cccc}
-1 & 0 & 0 & 0	\\
0 & 1 & 0 & 0	\\
0 & 0 & 1 & 0	\\
0 & 0 & 0 & 1
\end{array}\right] + a\left[\begin{array}{cccc}
0 & -\frac{p^{1}}{2} & -\frac{p^{2}}{2} & -\frac{p^{3}}{2}	\\
0 & -p^{0} & 0 & 0	\\
0 & 0 & -p^{0} & 0	\\
0 & 0 & 0 & -p^{0}
\end{array}\right] \label{A3}
\eea 
Thus, we find the action for the scalar field theory in $\kappa$-deformed Minkowski space-time is
\be
 \mathcal{S}_{\kappa-Minkowski}=\int d^{4}x \sqrt{-\hat{\eta}}L(\phi,\partial_{\mu}\phi,a), \label{A4}
\ee
where $\hat{\eta}=det~\hat{\eta}_{\mu\nu}=-(1-3ap^{0})$ and $L(\phi,\partial_{\mu}\phi,a)$ is given in Eq.(\ref{A1}). Next by varying the above action, we find the equation of motion  as
\be
\bigg\{-\partial_{0}^{2}+\partial_{x}^{2}+\partial_{y}^{2}+\partial_{z}^{2} -V(x)+ap^{0}\Big(\partial_{x}^{2}+\partial_{y}^{2}+\partial_{z}^{2}\Big)-\frac{a}{2}\Big(p^{1}\partial_{0}\partial_{x}+p^{2}\partial_{0}\partial_{y}+p^{3}\partial_{0}\partial_{z}\Big)\bigg\} \phi =0. \label{A5}
\ee
We take an infinitesimal variation of the action given in Eq.(\ref{A4})  and find the expression of symmetrized energy-momentum tensor as
\be
 \hat{T}_{\mu\nu}(a)= \partial_{\mu}\phi \partial_{\mu}\phi +\frac{1}{2}(\hat{\eta}_{\mu\nu}+\hat{\eta}_{\nu\mu})L(\phi,\partial_{\mu}\phi, a) \label{A6}
\ee
The Green's function corresponding to the Eq.(\ref{A5}) satisfy
\begin{multline}
\bigg\{-\partial_{0}^{2}+\partial_{x}^{2}+\partial_{y}^{2}+\partial_{z}^{2} -\lambda_{1}\delta(z-z_{1})-\lambda_{2}\delta(z-z_{2})+ap^{0}\Big(\partial_{x}^{2}+\partial_{y}^{2}+\partial_{z}^{2}\Big)\\-\frac{a}{2}\Big(p^{1}\partial_{0}\partial_{x}+p^{2}\partial_{0}\partial_{y}+p^{3}\partial_{0}\partial_{z}\Big)\bigg\}G(x,x^{\prime},a)=(1+\frac{3}{2}ap^{0})\delta(t-t^{\prime})\delta(z-z^{\prime})\delta(x_{\perp}-x_{\perp}^{\prime}), \label{A7}
\end{multline}
where $x_{\perp}=x,y$. Note that correction terms on the right-hand side are coming from expression $\sqrt{-\hat{\eta}}=(1-\frac{3}{2}ap^{0})$. We see that $V(x)=\lambda_{1}\delta(z-z_{1})+\lambda_{2}\delta(z-z_{2})$ has only $z$ dependency. Thus the Fourier transformation of Green's function is expressed as 
\be
G(x,x^{\prime},a)=\int_{\infty}^{\infty} \frac{d\omega d^{2}k_{\perp}}{8\pi^{3}}e^{-i\omega(t-t^{\prime})}e^{ik_{\perp}(x_{\perp}-x_{\perp}^{\prime})}g(z,z^{\prime},a), \label{A8}
\ee
where $g(z,z^{\prime},a)$ is the reduced Green's function.
Using Eq.(\ref{A7}) and Eq.(\ref{A8}) we find that the reduced Green's functions satisfy
\begin{multline}
-\bigg\{\partial_{z}^{2}+\omega^{2}-k_{\perp}^{2}-\lambda_{1}\delta(z-z_{1})-\lambda_{2}\delta(z-z_{2})+ ap^{0}\Big(\partial_{z}^{2}-k_{\perp}^{2}\Big)\\-a\Big(\frac{p^{1}}{2}(\omega k_{x})+\frac{p^{2}}{2}(\omega k_{y})+\frac{p^{3}}{2}(-i \omega)\partial_{z}\Big)\bigg\} g(z,z^{\prime},a)=(1+\frac{3}{2}ap^{0})\delta(z-z^{\prime}). \label{A9}
\end{multline}
Next by following the same procedure discussed in Section 5,  we perturbatively solve the above equation with $\omega \rightarrow i\zeta$ and find 
\bea
g(z,z^{\prime},a)&=&(1+\frac{ap^{0}}{2})g^{(0)}(z,z^{\prime})+ap^{0}\Bigg[\int d\bar{z}~g^{(0)}(z,\bar{z})\bigg\{\zeta^{2}+\lambda_{1}\delta(\bar{z}-z_{1})+\lambda_{2}(\bar{z}-z_{2})\nonumber\\&&-\frac{1}{2 p^{0}}\Big(p^{1}(\omega k_{x})+p^{2}(\omega k_{y})+p^{3}(-i \omega)\partial_{\bar{z}}\Big)\bigg\}g^{(0)}(\bar{z},z^{\prime})\Bigg], \label{A10}
\eea
where  
\bea
g^{(0)}(z,z^{\prime})&=& \Bigg[\frac{1}{2\bar{k}}e^{-\bar{k}|z-z^{\prime}|}-\frac{1}{2\bar{k}\Delta}e^{\bar{k}(z+z^{\prime}-2z_{1})}\bigg\{\frac{\lambda_{1}}{2\bar{k}}\Big(1+\frac{\lambda_{2}}{2\bar{k}}\Big)e^{2\bar{k}L} + \frac{\lambda_{2}}{2\bar{k}}\Big(1-\frac{\lambda_{1}}{2\bar{k}}\Big)\bigg\}  \Bigg],~~~where~\left\lbrace z, z^{\prime}\right\rbrace <z_{1}. \nonumber \\\label{A11}\\
&=& \Bigg[\frac{1}{2\bar{k}}e^{-\bar{k}|z-z^{\prime}|}-\frac{1}{2\bar{k}\Delta}\bigg\{\frac{\lambda_{1}}{2\bar{k}}\Big(1+\frac{\lambda_{2}}{2\bar{k}}\Big)e^{-\bar{k}(z+z^{\prime}-2z_{2})} + \frac{\lambda_{2}}{2\bar{k}}\Big(1+\frac{\lambda_{1}}{2\bar{k}}\Big)e^{\bar{k}(z+z^{\prime}-2z_{1})}\nonumber \\
&&~~~~ -\frac{\lambda_{1}\lambda_{2}}{4\bar{k}^{2}}2\cosh \bar{k}(z-z^{\prime})\bigg\}  \Bigg], ~~~~~~~~~~~~~~~~~~~~~~~~~~~~~~~~~~~~~~~~~~~~~~~~where~~z_{1}<\left\lbrace z, z^{\prime}\right\rbrace <z_{2}. \nonumber \\ \label{A12}\\
&=& \Bigg[\frac{1}{2\bar{k}}e^{-\bar{k}|z-z^{\prime}|}-\frac{1}{2\bar{k}\Delta}e^{-\bar{k}(z+z^{\prime}-2z_{2})}\bigg\{\frac{\lambda_{1}}{2\bar{k}}\Big(1-\frac{\lambda_{2}}{2\bar{k}}\Big) + \frac{\lambda_{2}}{2\bar{k}}\Big(1+\frac{\lambda_{1}}{2\bar{k}}\Big)e^{2\bar{k}L}\bigg\}  \Bigg],~where~\left\lbrace z, z^{\prime}\right\rbrace > z_{2}. \nonumber \\ \label{A13}
\eea
In the above expressions $\bar{k}=\sqrt{k_{\perp}^{2}+\zeta^{2}}$, $L=z_{2}-z_{1}$ and
\be
\Delta=\Big(1+\frac{\lambda_{1}}{2\bar{k}}\Big)\Big(1+\frac{\lambda_{2}}{2\bar{k}}\Big)e^{2\bar{k}L}-\frac{\lambda_{1} \lambda_{2}}{4\bar{k}^{2}}. \label{A14} 
\ee
Note that as in the previous cases (Eq.(\ref{L24}) for single plate and Eq.(\ref{L34}) for two parallel plates) here also the integral in the second term in Eq.(\ref{A10}) has limit either $z \rightarrow z^{\prime}$ for $z^{\prime}> z$ or $z^{\prime} \rightarrow z$ when $z > z^{\prime}$. Thus, only the first term will contribute in finding 
$g(z,z, a)$ from the above expression. In the later part of the discussion, we will see that to find the total energy associated with the plates, we only need $g^{(0)}(z,z)$. 

Next, we find the total energy for the two parallel plates kept at $z=z_{1}$ and $z=z_{2}$ when gravity is absent. For this using Eq.(\ref{A5}) in Eq.(\ref{A6}) first we calculate the $\hat{T}_{00}$ component as
\be
\hat{T}_{00}=\frac{1}{2}\partial_{t}\phi \partial_{t}\phi -\frac{1}{2}\phi \partial_{t}^{2}\phi +\frac{1}{4}\partial_{i}^{2}(\phi^{2})+\frac{ap^{0}}{4}\partial_{i}^{2}(\phi^{2})-\frac{a}{8}\bigg\{ p^{1}\partial_{t}\Big(\partial_{x}(\phi^{2})\Big)+p^{2}\partial_{t}\Big(\partial_{y}(\phi^{2})\Big)+p^{1}\partial_{t}\Big(\partial_{z}(\phi^{2})\Big)\bigg\}.\label{A15}
\ee
By taking vacuum expectation value of the above equation using $<\phi(x)\phi(x^{\prime})>=\frac{1}{i}G(x,x^{\prime})$ we find

\begin{multline}
<\hat{T}_{00}>=\frac{1}{2i}\Big\{(\partial_{t}\partial_{t^{\prime}}-\partial_{t}^{2})G(x,x^{\prime},a)\Big\}\Big\vert_{x=x^{\prime}}+\frac{1}{4i}\partial_{i}^{2}\Big\{G(x,x^{\prime},a)\Big\vert_{x=x^{\prime}}\Big\} + \frac{ap^{0}}{4i}\partial_{i}^{2}\Big\{G(x,x^{\prime},a)\Big\vert_{x=x^{\prime}}\Big\}\\-\frac{a}{8}\Bigg[p^{1}\partial_{t}\partial_{x}\Big\{G(x,x^{\prime},a)\Big\vert_{x=x^{\prime}}\Big\}+p^{2}\partial_{t}\partial_{y}\Big\{G(x,x^{\prime},a)\Big\vert_{x=x^{\prime}}\Big\}+p^{3}\partial_{t}\partial_{z}\Big\{G(x,x^{\prime},a)\Big\vert_{x=x^{\prime}}\Big\}\Bigg]. \label{A16}
\end{multline}
Next using Eq.(\ref{A8}) and Eq.(\ref{A16}) we find the total energy per unit area for the two parallel plates as 
\be
E_{tot}=\int \frac{dz d \omega d^{2}k_{\perp}}{8 \pi^{3}} \Big(1-\frac{3}{2}ap^{0}\Big) \Bigg[\frac{1}{2i}\Big(2\omega^{2} g(z,z,a)\Big)+\frac{1}{4i}\partial_{i}^{2}\Big(g(z,z,a)\Big) +\frac{ap^{0}}{4i}\partial_{i}^{2}\Big(g(z,z,a))\Big)\Bigg]. \label{A17}
\ee
Next, using the Gauss-divergence theorem and Eq.(\ref{A10}) with $\omega=i \zeta$ in the above equation we find the total energy per unit area, valid up to the first order in deformation parameter $a$ as
\be
E_{tot}=(1-ap^{0})\Bigg[-\frac{1}{2}\int \frac{d \zeta d^{2}k_{\perp}}{8 \pi^{3}}2\zeta^{2}\int^{+\infty}_{-\infty}dz~g^{(0)}(z,z)\Bigg]. \label{A18}
\ee
To evaluate the above expression using Eq.(\ref{A11}), Eq.(\ref{A12}) and Eq.(\ref{A13}) we first obtain 
 \be
 2 \zeta^{2} \int^{+\infty}_{-\infty}dz~g^{(0)}(z,z)=\frac{\zeta^{2}}{\bar{k}^{2}}\int^{+\infty}_{-\infty} \bar{k}dz -\frac{\zeta^{2}}{\bar{k}^{2}}\frac{1}{\Delta_{0}}\Bigg[\frac{\lambda_{1}}{2\bar{k}}+\frac{\lambda_{2}}{2\bar{k}}+2\frac{\lambda_{1}}{2\bar{k}}\frac{\lambda_{2}}{2\bar{k}}\bigg\{1-e^{-2\bar{k}L}\bigg\}-2\bar{k}L\frac{\lambda_{1}}{2\bar{k}}\frac{\lambda_{2}}{2\bar{k}}e^{-2\bar{k}L}\Bigg], \label{A19}
 \ee
where $\Delta_{0}=e^{-2\bar{k}L}\Delta$. Next using spherical polar coordinate for the phase space (see discussion after Eq.(\ref{L59})) and using the above equation in Eq.(\ref{A18}) we find 
%\footnote{ Here we use $\int^{+ \infty}_{-\infty} \frac{d \zeta d^{2}k_{\perp}}{16 \pi^{3}}\frac{\zeta^{2}}{\bar{k}^{2}} = \frac{1}{12\pi^{2}}\int^{\infty}_{0} \bar{k}^{2} d\bar{k}$.} 
\begin{multline}
 E_{tot}=-\frac{(1-ap^{0})}{12 \pi^{2}}\int^{\infty}_{0} \bar{k}^{3}d\bar{k}\int^{+\infty}_{-\infty}dz \\+\frac{(1-ap^{0})}{12 \pi^{2}}\int^{\infty}_{0}\frac{\bar{k}^{2}d\bar{k}}{\Delta_{0}}\Bigg[\frac{\lambda_{1}}{2\bar{k}}+\frac{\lambda_{2}}{2\bar{k}}+2\frac{\lambda_{1}}{2\bar{k}}\frac{\lambda_{2}}{2\bar{k}}\bigg\{1-e^{-2\bar{k}L}\bigg\}-2\bar{k}L\frac{\lambda_{1}}{2\bar{k}}\frac{\lambda_{2}}{2\bar{k}}e^{-2\bar{k}L}\Bigg]. \label{A20}
\end{multline}
Note here that the first term in the above equation is a divergent, which we define as bulk energy per unit area  ($\tilde{E}_{bulk}$) of the system. After some calculational simplification, we find the total energy per unit area associated with the two parallel plates in $\kappa$-deformed space-time as
\be
 E_{tot}=\tilde{E}_{bulk}+\tilde{E}_{d1}+\tilde{E}_{d2}+\tilde{E}_{c}, \label{A21}
\ee
where 
\be
\tilde{E}_{d1} = \frac{(1-ap^{0})}{96\pi^{2} L^{3}}\int^{\infty}_{0}dY Y^{2}\frac{1}{\Big(1+\frac{Y}{\lambda_{1}L}\Big)}~~;~~~~~ \tilde{E}_{d2} = \frac{(1-ap^{0})}{96\pi^{2} L^{3}}\int^{\infty}_{0}dY Y^{2}\frac{1}{\Big(1+\frac{Y}{\lambda_{2}L}\Big)}  \label{A22}
\ee
and 
\be
\tilde{E}_{c} = -(1-ap^{0})\frac{1}{96\pi^{2} L^{3}}\int^{\infty}_{0}dY Y^{3}\frac{1+\frac{1}{Y+\lambda_{1}L} +\frac{1}{Y+\lambda_{2}L}}{\Big(1+\frac{Y}{\lambda_{1}L}\Big)\Big(1+\frac{Y}{\lambda_{2}L}\Big)e^{Y}-1}.\label{A23}
\ee
Here $Y=2\bar{k}L$. Note that here $\tilde{E}_{d1}$, $ \tilde{E}_{d2}$ are self-energies of the two plates, and $\tilde{E}_{c}$ is the Casimir energy per unit area between the two parallel plates. Note here that we have included terms up to the first order in deformation parameter $a$. In the strong interation limit i.e., $\lambda_{1},\lambda_{2} \rightarrow \infty$ from expression of $\tilde{E}_{c}$ we find the $\kappa$-defomed Casimir energy for the two parallel plates. In the limit $a \rightarrow 0$, this reduces to the commutative result \cite{Mil}.

\subsection*{\bf Acknowledgments}
S.K.P thanks UGC, India, for the support through the JRF scheme (id.191620059604).

\end{document}